\documentclass[%
 reprint,
superscriptaddress,
 amsmath,amssymb,
aps,prl,
floatfix
]{revtex4-2}

\usepackage{graphicx}% Include figure files
\usepackage{dcolumn}% Align table columns on decimal point
\usepackage{bm}% bold math
\usepackage{color}

\usepackage{bbm}
\usepackage[colorlinks=true,bookmarks=false,allcolors=blue]{hyperref}
\hypersetup{pdfstartview={FitH},pdfpagemode={UseNone}, breaklinks=true,
            colorlinks,allcolors=blue, bookmarksopen=true, pdfnewwindow=true}
\usepackage[all]{hypcap}

\graphicspath{{pict/}{figures/}{}}

\usepackage[english]{babel}

\begin{document}

%\preprint{APS/123-QED}

\title{Metasurface-Assisted Quantum Ghost Discrimination of Polarization Objects}

%\title{Nonlocal Quantum Discrimination of Polarization Objects using Metasurfaces}
%-Sensitive 
%\title{Non-local Identification of Polarization-sensitive Objects using Metasurfaces}

\author{Andres Vega}%
 \email{andres.vega@uni-jena.de}
 \affiliation{
Institute of Applied Physics, Abbe Center of Photonics, Friedrich Schiller University Jena, Albert-Einstein-Str. 15, 07745 Jena, Germany}%
% \author{Kai Wang}
% \affiliation{
% Nonlinear Physics Centre, Research School of Physics, Australian National University, Canberra ACT 2601, Australia
% }%
\author{Thomas Pertsch}
\affiliation{
Institute of Applied Physics, Abbe Center of Photonics, Friedrich Schiller University Jena, Albert-Einstein-Str. 15, 07745 Jena, Germany}%
\affiliation{
Fraunhofer Institute for Applied Optics and Precision Engineering IOF, Albert-Einstein-Str. 7, 07745 Jena, Germany}
\author{Frank Setzpfandt}
\affiliation{
Institute of Applied Physics, Abbe Center of Photonics, Friedrich Schiller University Jena, Albert-Einstein-Str. 15, 07745 Jena, Germany}%
\author{Andrey A. Sukhorukov}
\email{andrey.sukhorukov@anu.edu.au}
\affiliation{
Research School of Physics, Australian National University, Canberra ACT 2601, Australia
}%
\affiliation{ARC Centre of Excellence for Transformative Meta-Optical Systems (TMOS),
Australia}

\date{\today}% It is always \today, today,

% ---------Added commands--------
\renewcommand{\P}{{\rm P}}
\newcommand{\R}{{\rm R}}
\renewcommand{\H}{{\rm H}}
\newcommand{\V}{{\rm V}}
\newcommand{\D}{{\rm D}}
\newcommand{\C}{{\rm C}}
\newcommand{\bra}[1] {\left\langle #1 \right|}
\newcommand{\ket}[1] {\left| #1 \right\rangle}
\renewcommand{\vector}[1] {\mathbf{#1}}
\newcommand\norm[1]{\left\lVert#1\right\rVert_F}
% -------------------------------

\begin{abstract}
We develop a concept of metasurface-assisted ghost imaging for non-local discrimination between a set of polarization objects.
The specially designed metasurfaces are incorporated in the imaging system to perform parallel state transformations in general elliptical bases of quantum-entangled or classically-correlated photons. 
Then, only four or fewer correlation measurements between multiple metasurface outputs and a simple polarization-insensitive bucket detector after the object can allow for 
%Our approach does not require any reconfigurable components, while allowing for 
the identification of fully or partially transparent polarization elements and their arbitrary orientation angles.
%without any reconfigurable components
We rigorously establish that entangled photon states offer a fundamental advantage compared to classical correlations for a broad class of objects. 
%This is achieved through only four or fewer parallel correlation measurements using specially designed metasurfaces performing photon state transformations in general elliptical bases.
%We prove that ghost imaging with quantum-entangled photons offers a fundamental advantage in polarization object discrimination, compared to the use of classically polarization-correlated light. 
The approach can find applications for real-time and low-light imaging across diverse spectral regions in dynamic environments.
%requiring only a simple polarization-insensitive bucket detector after the object.
%This only requires 
%operates with polarization-entangled photons, where one photon passes through and object 
%where 
%specially the metasurfaces We show that the metasurfaces can be tailored to transform the and separate 
%propose a non-local measurement scheme with polarization-entangled photons, where optimal nanostructured dielectric metasurfaces assist the discrimination between 
%different polarization objects
%of a known set with different polarization characteristics, including fully or partially transparent elements. We demonstrate the advantage of using a source with a high degree of entanglement when discerning a particular set of objects.
\end{abstract}

\maketitle

Optical imaging of polarization properties provides a rich amount of otherwise hidden information, with applications spanning from microscopy~\cite{Jan:2015-4705:BOE} to monitoring from satellites~\cite{Puthukkudy:2020-5207:RAR}. While manipulation and measurement of polarization is conventionally performed using bulk optical elements~\cite{Chekhova:2021:PolarizationLight}, nanostructured metasurfaces allow the most flexible in-parallel polarization transformations for single-shot measurements~\cite{Martinez:2018-750:SCI}, real-time imaging with a camera~\cite{Rubin:2019-eaax1839:SCI}, quantum light manipulation and characterization~\cite{Stav:2018-1101:SCI, Wang:2018-1104:SCI, Georgi:2019-70:LSA, Altuzarra:2019-20101:PRA, Solntsev:2021-327:NPHOT}. 
%The identification of object's polarization characteristics with 

Fundamental and applied interest in polarization detection at low-light illumination and across broad spectral regions, for example for bio-sensing, motivate the development of polarization ghost imaging~\cite{Kellock:2014-55702:JOPT}. Such schemes draw on the underlying principle originally developed for spatial ghost imaging, where the photons passing through an object are registered with a simple bucket detector~\cite{Pittman:1995-3429:PRA, Valencia:2005-63601:PRL}, while their quantum or classically-correlated pairs can be conveniently imaged at a different wavelength selected for efficient high-resolution detection~\cite{Chan:2009-33808:PRA, Karmakar:2010-33845:PRA, Aspden:2015-1049:OPT}. 
%the correlation between the measurements on the photon-pair allows to obtain the image of the object \cite{Erkmen:2010-405:ADOP}.
The object is characterized through multiple coincidence or correlation measurements~\cite{Erkmen:2010-405:ADOP} that can deliver a better signal-to-noise ratio compared to classical imaging systems, and also enable imaging with a very low number of photons \cite{Brida:2010-227:NPHOT, Morris:2015-5913:NCOM}.
However, there remains a fundamental limitation of traditional ghost polarimetry approaches due to a need for multiple reconfigurable elements such as rotating waveplates~\cite{Hannonen:2016-4943:OL, Hannonen:2017-1360:JOSA, Janassek:2018-883:OL, Shi:2018-100:OLE, Chirkin:2018-115404:LPL, Hannonen:2020-714:JOSA, Rosskopf:2020-34062:PRAP, Magnitskiy:2020-3641:OL}. Yet, the unique capabilities of polarization control with metasurfaces towards  potential single-shot ghost imaging configurations remains largely untapped, so far limited to the incorporation of metasurfaces for hologram generation~\cite{Liu:2017-e1701477:SCA}.

In this Letter we present a novel concept of metasurface-assisted polarization ghost imaging. We show that by placing specially designed metasurfaces before the polarization-insensitive photon detectors, one can perform discrimination between fully or partially transparent polarization-sensitive objects within a defined set. Furthermore, in our scheme the orientation angle of each object can be simultaneously recognized from only four or fewer parallel correlation measurements. This can facilitate real-time identification of different samples for potential applications including microscopy, whereas full ghost polarimetry requires at least eight measurements with multiple time-consuming reconfigurations to determine all elements of a general Jones matrix form~\cite{Hannonen:2020-714:JOSA}.
We also note that only discrimination of non-birefringent objects with different transversely varying transmission profiles and fixed orientations was realized previously~\cite{Malik:2010-163602:PRL}. 

Furthermore, we prove in a general way that ghost imaging with quantum-entangled photons offers a fundamental advantage in polarization object discrimination, compared to the use of classically polarization-correlated light. 
Entanglement is the most prominent feature in quantum mechanics and has become the cornerstone of rising technologies such as quantum computing \cite{Slussarenko:2019-41303:APR,Huang:2020-180501:SCIS}, communication \cite{Gisin:2007-165:NPHOT} and metrology \cite{Giovannetti:2011-222:NPHOT, Pezze:2018-35005:RMP}. This curious property enables unique non-local measurements since information about the properties of one particle can be obtained by performing a measurement on its entangled partner.
Whereas both classical and quantum schemes can be used for  imaging of non-polarizing greyscale objects~\cite{Valencia:2005-63601:PRL, Shapiro:2015-10329:SRP, Bennink:2002-113601:PRL, Gatti:2004-93602:PRL}, this does not hold for transparent objects~\cite{Altuzarra:2019-20101:PRA}.
Importantly, the correlations between quantum-entangled photons can provide additional information on the phase differences between various polarization components, which might not be detected with classical light. We rigorously derive the object properties that entail the need for entanglement.

\begin{figure}[t]
\centering
\includegraphics[width=\columnwidth]{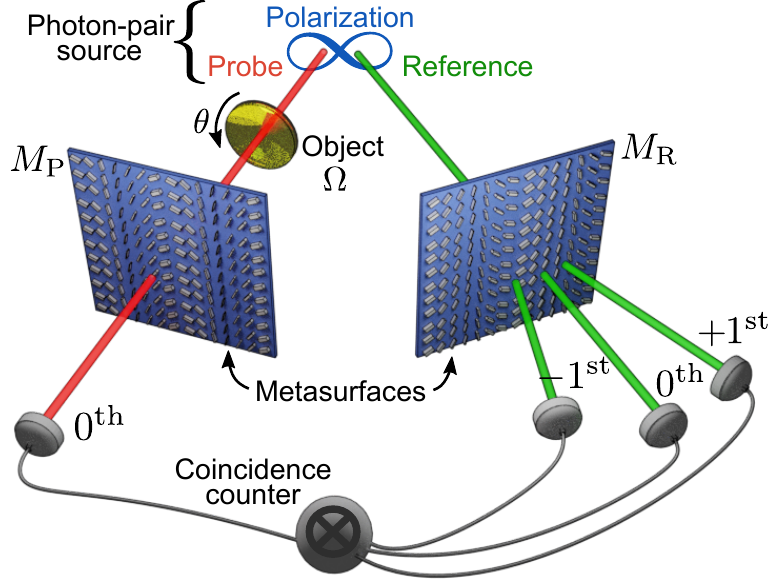}
\caption{\label{fig:setup} Sketch of the setup. Polarization transformations by metasurfaces in front of polarization-insensitive detectors enable discrimination of polarizaion objects $\Omega$ and their orientation angles $\theta$ through the coincidence measurements.}
\end{figure}

Our proposed measurement scheme is depicted in Fig.~\ref{fig:setup}. We consider a source producing a pair of probe and reference photons, which are entangled in the polarization state. Then, according to the principle of ghost imaging, only the probe photon passes through the object characterized by a polarization Jones matrix $\Omega$. We also consider a possible rotation of the object by an angle $\theta$, such that $\Omega(\theta) = R(\theta) \Omega R(-\theta)$, where $R(\theta)$ is a rotation matrix in the counterclockwise direction. The probe photon then passes through a metasurface and is registered by a polarization-insensitive click detector. The paired reference photon does not interact with the object. We place a tailored metasurface in its path, which splits the output between several polarization-insensitive detectors depending on the reference polarization state. We show in the following that by specially designing the metasurfaces, the coincidence measurements between the probe and reference at different detectors enables discrimination between polarization objects, and simultaneous identification of an arbitrary object orientation angle.

The polarization state of the photon-pairs
can be defined by the density matrix
\begin{eqnarray}
    \rho_{\mathrm{in}} &= \frac{1}{2}\left( \ket{\H_\P \H_\R} \bra{\H_\P \H_\R} + \ket{\V_\P \V_\R} \bra{\V_\P \V_\R}\right) \nonumber \\
     & + q\; \frac{1}{2} \left(\ket{\H_\P \H_\R} \bra{\V_\P \V_\R} + \ket{\V_\P \V_\R} \bra{\H_\P \H_\R}\right),
\end{eqnarray}
where $q$ is equal to the concurrence~\cite{Wootters:1998-2245:PRL} and represents the degree of entanglement with $0 \leq q \leq 1$, from the strongest classical correlation $q=0$ to perfect entanglement $q=1$, with the Bell parameter~\cite{Aspect:1982-91:PRL} value $S=\sqrt{2}(1+q)$. Throughout this manuscript, horizontal and vertical polarization are denoted H and V, respectively.
%
%The probe-photon interacts with an object $\Omega$, which furthermore can be rotated by an angle $\theta$, then is projected into a specific polarization state using the transformation $M_\P$, and finally detected. 
%
The transformation of the probe photon polarization is defined by the Jones polarization transfer matrices of the object $\Omega \left(\theta\right)$ and the metasurface $M_\P$. For the reference photon, there are several outputs after the metasurface with the Jones matrices $M_{\R,n}$ for the different diffraction orders $n$.

%The probe photon transformation after the object can be defined through a metasurface Jones matrix $M_\P$. Then,
After detection of the probe photon, the reduced state of the reference photon before the metasurface $M_R$ is determined according to the principle of remote state preparation~\cite{Peters:2005-150502:PRL} as $\rho_\R' = \mathrm{tr}_\P \left[ (T_\P \otimes \mathbbm{1}) \, \rho_\mathrm{in} \, (T_\P \otimes \mathbbm{1})^\dagger \right]$, where $\mathrm{tr}_\P(\cdot)$ is the partial trace over the probe-photon and $T_\P=M_\P \Omega \left(\theta\right)$. %is the total transformation that contains the object $\Omega$ and projector $M_\P$ that interact with the probe-photon.
The normalized reduced state
%of the reference-photon after measuring the probe-photon 
is then $\rho_\R = \rho_\R'/\mathrm{tr}(\rho_\R')$.
%The reference-photon does not interact with the object, but is projected onto several polarization bases by interacting with an optical element realizing the transformation $M_\R$. We assume, that each of these polarization bases corresponds to a separate output channel, in which the probe photons then are detected using polarization-insensitive detectors. Whereas the measurements of the probe and reference photons alone do not allow to distinguish different objects $\Omega$ and their rotation angle $\theta$, we will demonstrate that this is possible using coincidence measurements. 
The expectation values of the coincidence counts between the
probe and reference detectors are
%single output of projector $M_\P$ and the $n$-th output of projector $M_\R$ is 
$\Gamma_n = \textrm{tr} \left[ (M_{\R,n}) \rho_\R (M_{\R,n})^\dagger \right]$. We show in the following that 
%To distinguish objects, 
the metasurfaces can be optimized to realize such $M_\P$ and $M_{\R,n}$, 
%each projection basis $M_{\R,n}$ in the reference arm is numerically engineered so 
that every reduced state $\rho_\R$ produces a distinctive pattern formed by a collection of 
%$n$ 
coincidence measurements, thereby allowing discrimination between the objects. 
%We proceed to elaborate on each of the design stages.

%We will show that it allows to distinguish polarization-sensitive objects $\Omega$ and their rotation angle $\theta$. The scheme consists of a source of polarization-entangled photon-pairs named probe (P) and reference (R). The probe-photon interacts with an object $\Omega$ and then passes through an element called $M_\P$ that projects the incoming state onto only one polarization basis. On the other hand, the reference-photon does not interact with the object but it does with an object $M_\R$ that projects the state onto several polarization bases. Lastly, detectors insensitive to polarization measure the probe and reference photons after their respective projectors and the measurements go into a coincidence counter. The following theoretical model delves into the details of the optimal design of the projectors.

% ----- OVERVIEW OF THE MODEL------

%As an overview, the strategy to design the polarization projectors consists of two stages. First, the projector $M_\P$ in the probe arm is numerically optimized so that each object leads to a unique reduced state of the reference-photon after the probe-photon is measured $\rho_\R' = \mathrm{tr}_\P \left[ (T_\P \otimes \mathbbm{1}) \, \rho_\mathrm{in} \, (T_\P \otimes \mathbbm{1})^\dagger \right]$ where $\mathrm{tr}_\P(\cdot)$ is the partial trace over the probe-photon, $T_\P=M_\P \Omega $ is the total transformation that contains the object $\Omega$ and projector $M_\P$ that interact with the probe-photon. The normalized reduced state of the reference-photon after measuring the probe-photon is then $\rho_\R = \rho_\R'/\mathrm{tr}(\rho_\R')$. Second, the expectation value of the coincidence counts between the single output of projector $M_\P$ and the n-th output of projector $M_\R$ is $\Gamma_n = \textrm{tr} \left[ (M_{\R,n}) \rho_\R (M_{\R,n})^\dagger \right]$, therefore, each projection basis $M_{\R,n}$ in the reference arm is numerically engineered so that every reduced state $\rho_\R$ produces a distinctive pattern formed by a collection of $n$ coincidence measurements. We proceed to elaborate on each of the design stages.

%%-------- First stage -------
We determine the optimal choice of the $M_\P$ metasurface transformation for the probe photon 
by geometrically representing the reduced state of the reference photon $\rho_\R$ in the Poincar\'e sphere~\cite{Goldberg:2021-1:ADOP}, where each $\rho_\R$ is described by a vector $\vector{p}=[p_\H, p_\D, p_\C]$. Here, $p_\H$ corresponds to the degree of horizontal or vertical polarization, $p_\D$ to the diagonal [$(\ket{H} + \ket{V})/\sqrt{2}$] or antidiagonal linear polarization at $\pm 45^\circ$, and $p_\C$ to the right [$(\ket{H} - i\ket{V})/\sqrt{2}$] or left circular polarization. The Poincar\'e vector is found using the Pauli matrices $X, Y, Z$ with $\vector{p}=\left[\mathrm{tr}(\rho_\R Z), \mathrm{tr}(\rho_\R X), -\mathrm{tr}(\rho_\R Y) \right]$. If the total transformation in the probe-arm, comprising the object $\Omega$ and metasurface $M_\P$, is written in the basis of the photon-pair source as $T_\P = \sum T_{kl}\ket{k} \bra{l}$ with $k,l \in \{\H, \V \}$, then
\begin{eqnarray} \label{eq:p1}
    p_\H &= &\norm{T_\P}^{-2}\left( |T_{\H \H}|^2+|T_{\V \H}|^2 - |T_{\V \V}|^2-|T_{\H \V}|^2 \right),\\ \label{eq:p2}
    p_\D &= &q\, \norm{T_\P}^{-2}\mathrm{Re} \left( T_{\H \H} T_{\H \V}^\ast + T_{\V \H} T_{\V \V}^\ast \right) ,\\ \label{eq:p3}
    p_\C &= &q\, \norm{T_\P}^{-2}\mathrm{Im} \left( T_{\H \H} T_{\H \V}^\ast + T_{\V \H} T_{\V \V}^\ast \right),
\end{eqnarray}
where $\norm{\cdot}$ is the Frobenius norm. Since we require that each object produces a different reduced state, each $\rho_\R$ must occupy a unique position on or inside the Poincar\'e sphere.

We reveal an important relation between the components of the Poincar\'e vector as
\begin{equation}
    p_\H^2 + \frac{p_\D^2 + p_\C^2}{q^2} = \eta^2 = \left( \frac{\sigma_{1\P}^2 - \sigma_{2\P}^2 }{\sigma_{1\P}^2 + \sigma_{2\P}^2}\right)^2,
    \label{eq:ellipsoid}
\end{equation}
where $\sigma_{1\P}$ and $\sigma_{2\P}$ are the singular values of the transformation $T_\P$, see Supplementary~S1 for the derivation. Eq.~(\ref{eq:ellipsoid}) indicates that the reduced state of the reference-photon $\rho_\R$ obtained from the family of transformations $T_\P$ which have the same $\eta$, will lie on an ellipsoid of revolution of long axis $\eta$ and short axis $q \eta$ given that the degree of entanglement is in the range $0\leq q\leq 1$. This ellipsoid reduces to a point, $p_\H=p_\D=p_\C=0$, when $\sigma_{1\P} = \sigma_{2\P}$. Additionally, the ellipsoid is largest, for a fixed $q$, when one of the singular values is zero and $\eta=1$.
%, $p_\H^2 + (p_\D^2 + p_\C^2)/q^2=1$.
For the moment, we concentrate on a scheme with a source of photon-pairs that have maximal entanglement $q=1$, where the ellipsoid is a sphere with radius $\eta$. 
%At the end of this letter we analyze the effect of photon-pairs with less entanglement, modeled with a smaller value of $q$, until the photon-pairs are only classically correlated.

We now use the formulated properties of the photon states 
%ground to understand the first stage of the model, we use it 
to understand the need of the transformation $M_\P$ in the probe arm for object identification. To this end, we illustrate the model with a set of three objects: a transparent object (quarter-wave plate) $\Omega_a = \ket{\H}\bra{\H} + \exp{(i\pi/2)}\ket{\V}\bra{\V}$  with singular values $\sigma_{1\Omega}= \sigma_{2\Omega}= 1$, a partially transparent object $\Omega_b = \ket{\H}\bra{\H} + \exp{[i(\pi/2 +0.7i)]}\ket{\V}\bra{\V}$ with $\sigma_{1\Omega}= 1$ and $\sigma_{2\Omega}= 0.5$, and a fully polarizing object (horizontal polarizer) $\Omega_c = \ket{\H}\bra{\H}$ with $\sigma_{1\Omega}= 1$ and $\sigma_{2\Omega}= 0$. These definitions are for an object rotation angle $\theta=0$.
%, measured at the horizontal polarization. 
%Given that we are also interested in identifying the rotation of each of the objects, bare in mind that
Since rotation is a unitary transformation, the mentioned singular values of each object remain unaffected and the associated reference-photon reduced states lie on a sphere with a radius defined by Eq.~(\ref{eq:ellipsoid}) for all object angles. Since the considered objects have a rotation symmetry of $\pi$, we aim to have different $\rho_\R$ in the interval $0\le \theta<\pi$ for the angle identification. 
%makes a round-trip after a rotation of $\pi$. 
%hence their reduced states create closed loops where each lies on a sphere with a radius defined by their singular values, as described by Eq.~(\ref{eq:ellipsoid}).

\begin{figure}[t]
\centering
\includegraphics[width=\columnwidth]{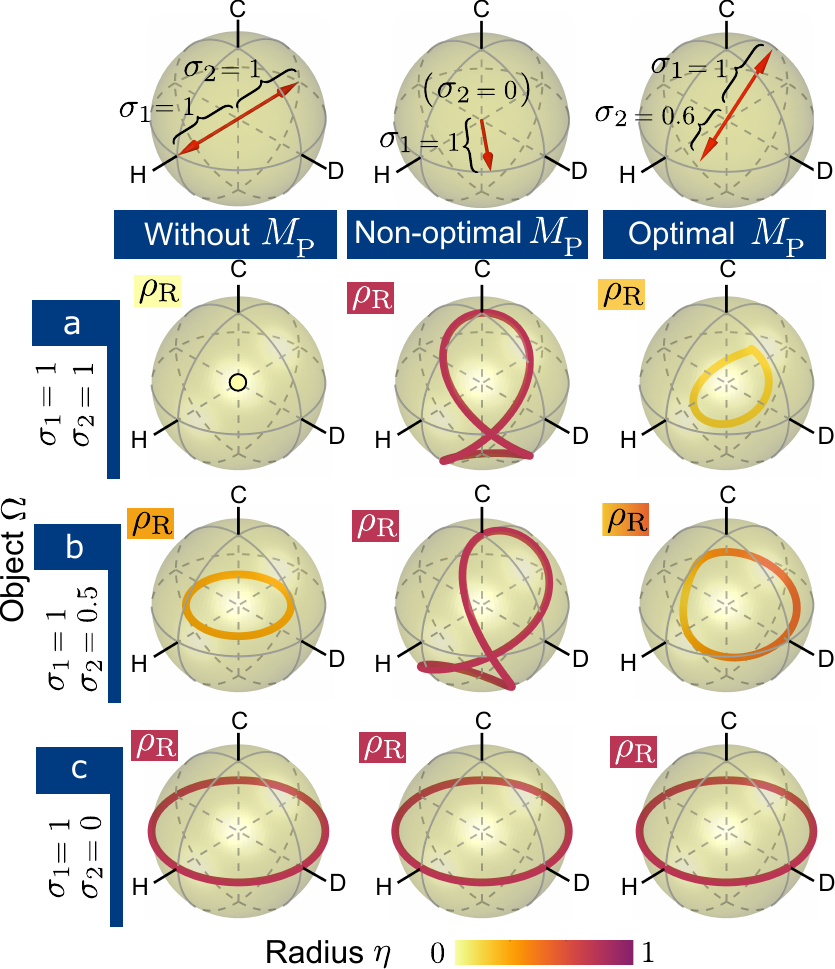}
\caption{\label{fig:MP} 
Metasurface transformation $M_\P$ enabling the object discrimination.
Top: The singular eigenvectors of $M_\P$, scaled by the corresponding singular values, represented in the Poincar\'e sphere for the cases of (left) no metasurface, (middle) fully polarizing transformation, (right) optimal partially polarizing case. The axes of the Poincar\'e sphere correspond to horizontal (H), diagonal (D) and right circular (C) polarizations.
(a-c)~Reduced states of the reference-photon after measuring the probe-photon $\rho_\R$ produced by a set of three objects $\Omega_{a,b,c}$, for different $M_\P$ in each column. The curves are formed from the points corresponding to different object rotation angles in the range $0\le\theta<\pi$.}
%with and without a projector $M_\P$ located in the probe arm.}
\end{figure}

We illustrate the effect of metasurface transformation $M_\P$ on the discrimination of objects in Fig.~\ref{fig:MP}. We compare the cases of no metasurface with $M_\P = I$ in the left, non-optimal transformation in the middle, and optimized transformation in the right column. In the top row, we show in the Poincar\'e sphere the two orthonormal right singular vectors of $M_\P$, $\vector{m}_{j\P}$ with $j \in \{1,2\}$ (see Supplementary~S2), whose magnitude has been scaled with their corresponding singular value $\sigma_{j\P}$. The rows (a-c) represent the normalized reduced states of the probe photon for the three chosen objects $\Omega_{a,b,c}$, respectively. 

For the simplest scenario without a metasurface $M_\P$ in the probe arm presented in Fig.~\ref{fig:MP}(left column), a fully transparent wave-plate $\Omega_a$ always corresponds to a single point,
%in agreement with Eq.~(\ref{eq:ellipsoid}), 
such that its orientation angle cannot be identified. 
Indeed, in this case $T_\P=\Omega$ with singular values $\sigma_{\P}=\sigma_{\Omega}$, and according to Eq.~(\ref{eq:ellipsoid}) any transparent phase object with $\sigma_{1\Omega}= \sigma_{2\Omega}$ will generate reduced states $\rho_\R$ that stay on the origin.
Hence, discrimination between transparent phase objects and determination of their rotations cannot be performed without a metasurface $M_\P$ acting on the polarization of the probe photon before a polarization-insensitive detector.
%, reduced states after detection of the probe photon are shown in the first column of Fig.~\ref{fig:MP}. %the probe-photon only interacts with an object and then it is measured by the detector.
%Here, $T_\P=\Omega$ with singular values $\sigma_{\P}=\sigma_{\Omega}$. According to Eq.~(\ref{eq:ellipsoid}), this implies that the phase object $\Omega_a$ with $\sigma_{1\Omega}= \sigma_{2\Omega}$ will generate reduced states $\rho_\R$ that stay on the origin at any rotation angle, while the other two objects with $\sigma_{1\Omega} \neq \sigma_{2\Omega}$ produce reduced states that stay out of the origin and their rotation can be easily identified. Hence, rotation of transparent phase objects cannot be identified without an adequate projection $M_\P$ acting on the polarization of the probe photon. %In the end, the mentioned impasse with the identification of the rotation of the phase object justifies the need for additional element in the probe arm, which should adequately act on the polarization state of the photon before reaching the detector.

%We see in Fig.~\ref{fig:MP}(left column) that without a metasurface, a fully transparent wave-plate $\Omega_a$ always corresponds to a central point in agreement with Eq.~(\ref{eq:ellipsoid}), such that its orientation angle cannot be identified. 

In Fig.~\ref{fig:MP}(middle column), we illustrate a case of fully polarizing transformation $M_\P$ with $\sigma_{2\P}=0$, such that only one polarization is transmitted. This represents a quantum projection measurement for the state $\vector{m}_{1\P}$. 
%Equation~(\ref{eq:ellipsoid}) can also be useful to understand the properties of the singular values of the projector $M_\P$. It is straightforward to show, that 
We note that if an object $\Omega$ and/or $M_\P$ have one singular value equal to zero, then its product, $T_\P=  M_\P \Omega$, also has one null singular value. 
%Physically, this means that only one polarization state is transmitted, i.e. the object represents a polarizer.
%This situation is illustrated in the last row of Fig.~\ref{fig:MP} where $\Omega$ has one null singular value and similarly for $M_\P$ in the second column. 
This causes, according to Eq.~(\ref{eq:ellipsoid}), the reduced state of the reference-photon to lie on the surface of the Poincar\'e sphere. Consequently, 
%if the additional element $M_\P$ has one null singular value, 
not all objects and rotation angles can be distinguished since the closed loops formed by the objects' reduced states would cross. 

To enable discrimination in the general case, the ratio of the smaller to larger singular values of $M_\P$ has to be in the range $0 <\sigma_2/\sigma_1 < 1$, i.e., it should be a partially polarizing transformation. Additionally, its largest singular value should be close to 1 to ensure high photon counts. The reduced states for a numerically optimized polarization basis of $M_\P$ are shown in Fig.~\ref{fig:MP}(right column). They do not cross and enable discrimination, confirming the above analysis of the required singular values. The displayed elliptical polarization basis is formed by the two orthonormal right singular vectors of $M_\P$, represented by arrows in the top-row plot.
%Poincar\'e sphere by $\vector{m}_{j\P}$ with $j \in \{1,2\}$ \cite{Note1}, whose magnitude has been scaled with their corresponding singular value $\sigma_{j\P}$. 
Importantly, the optimal $M_\P$ depends on the given set of objects.
%; however, $M_\P$ may not be necessary if all the reduced states $\rho_\R$ of a set of objects are already separated from each other in the Poincar\'e sphere.

%% -------------Second stage-----------------------
Next, we discuss conditions on the optimal polarization bases of the transformation $M_\R$ in the reference arm, which ensures that each object $\Omega$ has a unique pattern in the coincidences. To achieve practically relevant solutions, we impose two additional conditions: object identification must be achieved using a small number of outputs and the total number of photon counts in all outputs should be larger than zero for any object. We note that a specially designed metasurface with four diffraction outputs can facilitate a full characterization of single-photon polarization~\cite{Wang:2018-1104:SCI}, thereby allowing for object discrimination based on the different reference photon states $\rho_R$. We find that discrimination between particular sets of objects may be performed with a smaller set of outputs, which can simplify the measurements and improve signal-to-noise ratio since photon counts are concentrated in fewer outputs. 
%To investigate this possibility,
The optimal design that meets these requirements can be investigated by writing the expectation value of the coincidence counts with the $n$-th reference output as (see Supplementary~S2)
\begin{equation}
\begin{split}
    \Gamma_n =&  \tfrac{1}{2}\left( \sigma_{1\R,n}^2 + \sigma_{2\R,n}^2 \right) \\
                        & + \tfrac{1}{2}\left( \vector{p}\cdot\vector{m}_{1\R,n} \right)  \left( \sigma_{1\R,n}^2 - \sigma_{2\R,n}^2 \right),
\end{split}
\end{equation}
where $\sigma_{\R,n}$ stands for a singular value of the transformation $M_{\R,n}$, and $\vector{m}_{\R,n}$ is its right singular vector represented in the Poincar\'e sphere. Additionally, the output pattern in coincidences can be represented %, for the sake of convenience,
in a space $\mathcal{D}$ such that each expectation value $\Gamma_n$ corresponds to a coordinate along the $n$-th dimension and each pattern in coincidences is mapped as a point in $\mathcal{D}$.

One approach to fulfill the requirement of non-zero photon counts consists of having equal singular values for all outputs of the metasurface $M_{\R}$, $\sigma_{1\R,n}=\sigma_{1\R}$ and $\sigma_{2\R,n}=\sigma_{2\R}$, and also that the sum of their right singular vectors to be equal to zero, $\sum \vector{m}_{1\R,n} = 0$. Thus, $\sum \Gamma_n = (1/2) \sum \left(  \sigma_{1\R}^2 + \sigma_{2\R}^2 \right)$ is a constant different from zero, meaning that the photon counts will be non-zero. If we consider $M_\R$ with three outputs, the condition $\sum \vector{m}_{1\R,n} = 0$ would indicate that each pair of  vectors $\vector{m}$ encloses an angle of $120^\circ$ and they would all lie in one plane in the Poincar\'e sphere. This plane has to be chosen such that each object $\Omega$ has a unique pattern in the coincidences. To demonstrate this, we return to our example with objects $\Omega_{a,b,c}$.

\begin{figure}[htb]
\centering
\includegraphics[width=\columnwidth]{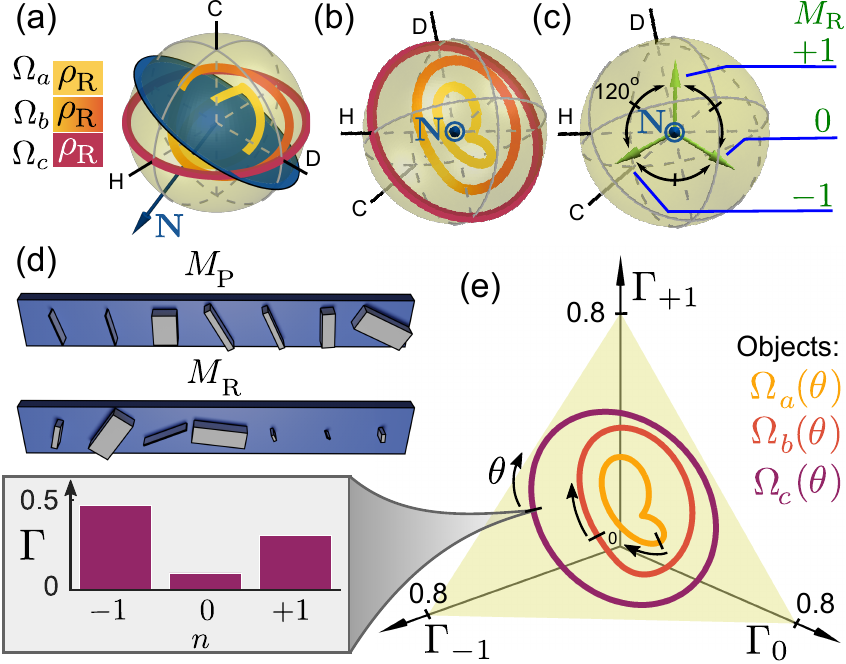}
\caption{\label{fig:MR_result} (a, b) Reduced state of the reference-photon $\rho_\R$ produced by a set of three objects $\Omega_{a,b,c}$ and an optimal metasurface $M_\P$ corresponding to Fig.~\ref{fig:MP}(right column) from two different perspectives. The vector $\vector{N}$ is normal to the blue plane where the optimal projection bases of the diffraction orders of the metasurface $M_\R$ are located, see (c). (d) Optimal meta-gratings in the probe $M_\P$ and reference $M_\R$ arms. (e) Coincidences at three diffraction orders $\Gamma_{-1}$, $\Gamma_{0}$, $\Gamma_{+1}$,
%for each of the objects $\Omega_{a,b,c}$ in the set, depicted in a space $\mathcal{D}$ where coincidences at each diffraction order is a dimension, 
showing that the objects can be discerned from one another and their own rotation angle $\theta$ unambiguously identified ($\theta=0$ is marked with ticks, the object rotation symmetry is $\pi$). The inset shows one of the coincidence patterns.}
\end{figure}

The reduced states $\rho_\R$ from the objects $\Omega\left(\theta\right)$ and the optimal $M_\P$ discussed in Fig.~\ref{fig:MP} are depicted together in the Poincar\'e sphere of Fig.~\ref{fig:MR_result}(a). A plane that can encompass the three vectors $\vector{m}$ while allowing for separation of the projected reduced states is illustrated in blue along with its normal $\vector{N}$. The projections of all the reduced states $\rho_\R$ into this plane are well separated, as shown in Fig.~\ref{fig:MR_result}(b). They can then be mapped into the space $\mathcal{D}$ of the coincidences at the outputs using the three projection vectors of $M_\R$ depicted in Fig.~\ref{fig:MR_result}(c), as described above. %These three vectors have the additional property that they map the geometric shape of this projection, Fig.~\ref{fig:MR_result}(a, middle), into the space $\mathcal{D}$ of the patterns in coincidences, as will be shown.
The direction of the optimal vector $\vector{N}$, and therefore the plane where the vectors $\vector{m}_{1\R,n}$ lie can be found numerically. %under the condition that the projected reduced density states $\rho_\R$ do not overlap [see Fig.~\ref{fig:MR_result}(a, right)].
For the set of objects investigated here, %Notice that the number of outputs of $M_\R$ has been numerically found,
at least three outputs are needed to achieve discrimination. %object identification for this particular example,
Two or four outputs may be needed for other sets of objects (see Supplementary~S5 for an example).

%------------ Implementation ---> Final result------------
The optimal form of $M_\P$ and $M_\R$ in each arm of the setup can be found numerically. %following the two-staged design that was presented;
%However, performing these polarization transformations experimentally can be quite challenging, as they in general need to realize partial polarizers in elliptical bases, and simultaneously for several outputs in the reference arm.
We note that the polarization transformations in principle can also be implemented with a collection of bulk optical elements; however, at the expense of complex designs sensitive to alignment.
%requiring large experimental space .
%to achieve the required polarizing bases.
%also needs several outputs. 
%As it was shown for the example set, $M_\P$ in the probe arm needs to realize a partially polarizing component in an elliptical base and $M_\R$ in the reference arm requires several outputs and also in elliptical bases.
Here we show that these polarization manipulations can be realized using 
%Our platform of choice to act on the polarization states of the photons in each arm are
nanostructured dielectric metasurfaces, which can effectively act as partial polarizers in arbitrary elliptical bases with any required extinction ratio \cite{Wang:2018-1104:SCI, Lung:2020-3015:ACSP}. Each metasurface is a flat optical element composed of a periodically repeated array of nano-resonators called meta-gratings. We perform optimization of the phase retardances and orientations associated with the individual nano-resonators to realize the required transformations, see details in Supplementary~S4 and~S6. Specifically, we design a metasurface for the probe arm such that its zeroth diffraction order realizes $M_\P$. Similarly, a second metasurface is placed in the reference arm where the three outputs of $M_\R$ correspond to three diffraction orders, as illustrated in Fig.~\ref{fig:setup}. The unit-cell geometries of numerically found optimal meta-gratings for the considered set of objects $\Omega_{a,b,c}$ are depicted in Fig.~\ref{fig:MR_result}(d). 
%We note that the polarization transformations in principle can also be implemented with a collection of bulk optical elements; however, at the expense of large experimental space and very complex designs to achieve the required polarizing bases.

The main result of this Letter is shown in Fig.~\ref{fig:MR_result}(e), where coincidence patterns at three diffraction orders for the objects $\Omega_{a,b,c}(\theta)$ are represented in the space $\mathcal{D}=\{\Gamma_{-1},\Gamma_{0},\Gamma_{+1}\}$, where each point on the plotted curves corresponds to a different object rotation angle $\theta$.
%particular distribution of the correlations across the three diffraction orders of the metasurface. 
The inset shows an example for the polarizing object $\Omega_c$ at $\theta = 0$. This showcases that the full identification of objects with their corresponding rotation angle is possible by optimizing $M_\P$ and $M_\R$. 
%All the coincidences correspond to practically relevant non-zero photon counts.
%since they lie on the plane $\Gamma_{-1} + \Gamma_{0} + \Gamma_{+1} \approx 0.8$ within space $\mathcal{D}$. 
As designed, the chosen three outputs of metasurface $M_\R$ map the geometric shape of Fig.~\ref{fig:MR_result}(b) into the space $\mathcal{D}$ in such a way that all the coincidences correspond to practically relevant non-zero photon counts.

%------------ENTANGLEMENT VS CLASSICAL CORRELATION - Example set: PHASE OBJECTS------------------------
After demonstrating that objects and their rotation angles can be discriminated using entangled states of light, next we show that entanglement is necessary to distinguish specific classes of objects. %Lastly, we demonstrate the advantage of entanglement to discern all the objects in a set.
To this end, we %relax the condition of a source with perfectly entangled photon-pairs, $q=1$, and
explore the effect of reducing the entanglement until reaching only classical correlation, $q=0$. We consider a different set of polarization-sensitive objects that contains only generalized retarders $\Omega = \ket{\H}\bra{\H} + \exp{(i\phi)}\ket{\V}\bra{\V}$, each with its own phase difference $\phi$ between the horizontal and vertical polarization. Here, we do not seek to identify the rotation of each of them but just discern them based on $\phi$.

\begin{figure}[t]
\centering
\includegraphics[width=\columnwidth]{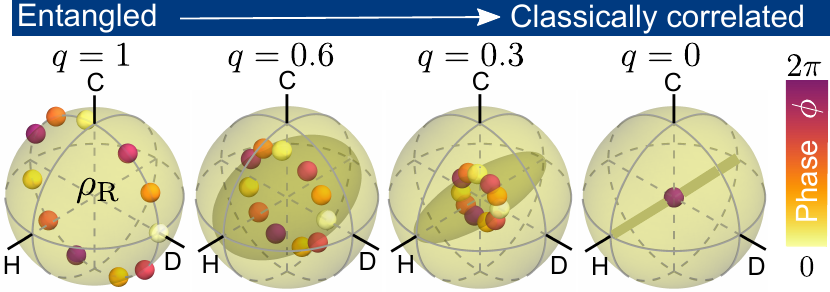}
\caption{\label{fig:q} Reduced state of the reference-photon after measuring the probe-photon $\rho_\R$ at different levels of entanglement. Here, the probe-photon interacts with a set of different fully transmissive phase objects, $\Omega = \ket{\H}\bra{\H} + \exp{(i\phi)}\ket{\V}\bra{\V}$ (phase $\phi$ shown in the color scale), one at a time, and also with an optimal $M_\P$ that acts as a diagonal polarizer.}
\end{figure}

%Following the analysis of the first stage of the method, an
The optimal $M_\P$ necessary to discern these phase objects for any value of $q$ %and it is numerically found to be
is a diagonal polarizer. Notice here that this element can obviously be implemented simply using a conventional polarizer instead of a metasurface. The corresponding reduced states $\rho_\R$ of this set of phase objects are depicted in Fig.~\ref{fig:q} for different levels of entanglement $q$. As can be seen, this chosen set of phase objects is a quite peculiar one because their reduced states have the same $p_\H$ Poincar\'e vector component [see Eq.~(\ref{eq:p1})], as $p_\H$ does not depend on the phase $\phi$. Furthermore, Eqs.~(\ref{eq:p2}) and (\ref{eq:p3}) show, that the lower the degree of entanglement, the smaller $p_\D$ and $p_\C$ become, as illustrated in Fig.~\ref{fig:q} with a shrinking ellipsoid according to Eq.~(\ref{eq:ellipsoid}). In the case of only classical correlation between photons with $q=0$, we have $p_\D=p_\C=0$, such that all reduced states are in the same position and the different objects cannot be discriminated. 
%the phase objects of the proposed set as they all lie in the same position.
Noteworthy, if the aforementioned phase objects are rotated by an angle $\theta$, their Poincar\'e component $p_\H$ becomes dependent on the phase $\phi$ and they could be identified using only classical correlations (see Supplementary~S3). In a nutshell, transformations $T_\P$ that have the same $p_\H$ component can only be discriminated if the photon-pair source has a high level of entanglement. Our conclusion on the advantage of entanglement generally agrees with the results of Ref.~\cite{Altuzarra:2019-20101:PRA}, where images of polarization-dependent patterns imprinted with a metasurface had higher visibility across different orientation angles for larger Bell parameter values.
%, where effectively a horizontal and a vertical polarizer rotated $45^\circ$ and no projector $M_\P$ were used.

%------------CONCLUSION-------------------------
In summary, we proposed a general approach for ghost discrimination between polarization-sensitive objects and simultaneous identification of their rotation through non-local measurements using non-classical light. This is achieved by using nanostructured metasurfaces designed to perform tailored polarization transformations.
%, harness modern advanced nanofabrication capabilities for high-fidelity operation. 
Importantly, we also proved that non-classical correlations between probe and reference-photons are needed to enable discrimination of all types of objects. 
%Our proposed measurement scheme can be implemented with dielectric metasurfaces that harness modern advanced nanofabrication capabilities for high-fidelity operation. 
We believe that these fundamental results can stimulate the practical development of 
%approach can satisfy the strong demand in research and industry for 
efficient and integrated optical schemes for characterization of objects with polarization-sensitive transmission characteristics across a broad spectral range. In particular, these can benefit applications demanding accurate discrimination of objects with various polarization characteristics, including sets of phase objects like biological samples.

%\textcolor{red}{Additionally, there is a strong demand in research and industry for efficient and integrated optical schemes performing characterization of objects with polarization-sensitive transmission characteristics across a broad spectral range. Nanostructured metasurfaces have demonstrated unprecedented flexibility in detection and imaging of polarization states of classical \cite{Yu:2014-139:NMAT, Rubin:2019-eaax1839:SCI} and quantum light \cite{Wang:2018-1104:SCI,Lung:2020-3015:ACSP}, offering an integrated and more capable alternative to conventional bulk optical systems.}

%-------------ACKNOWLEDGMENTS------------------
\begin{acknowledgments}
The authors thank Michael Brodsky, Kai Wang, Shaun Lung, Jihua Zhang for insightful discussions and comments.
This work was supported by the Thuringian Ministry for Economy, Science, and Digital Society (2021 FGI 0043), the German Federal Ministry of Education and Research (FKZ 13N14877), the German Research Foundation (IRTG 2675), the European Union through the ERASMUS+ program, the German Academic Exchange Service (grant 57388353), and UA-DAAD exchange scheme. A.A.S. also acknowledges the support by the Australian Research Council (DP160100619, DP190101559, CE200100010) and US AOARD (19IOA053). 
%\textcolor{red}{@THOMAS: can we put IRTG already?}
\end{acknowledgments}

%\end{document}
\bibliography{db_ghost_meta}

\begin{thebibliography}{42}%
\makeatletter
\providecommand \@ifxundefined [1]{%
 \@ifx{#1\undefined}
}%
\providecommand \@ifnum [1]{%
 \ifnum #1\expandafter \@firstoftwo
 \else \expandafter \@secondoftwo
 \fi
}%
\providecommand \@ifx [1]{%
 \ifx #1\expandafter \@firstoftwo
 \else \expandafter \@secondoftwo
 \fi
}%
\providecommand \natexlab [1]{#1}%
\providecommand \enquote  [1]{``#1''}%
\providecommand \bibnamefont  [1]{#1}%
\providecommand \bibfnamefont [1]{#1}%
\providecommand \citenamefont [1]{#1}%
\providecommand \href@noop [0]{\@secondoftwo}%
\providecommand \href [0]{\begingroup \@sanitize@url \@href}%
\providecommand \@href[1]{\@@startlink{#1}\@@href}%
\providecommand \@@href[1]{\endgroup#1\@@endlink}%
\providecommand \@sanitize@url [0]{\catcode `\\12\catcode `\$12\catcode
  `\&12\catcode `\#12\catcode `\^12\catcode `\_12\catcode `\%12\relax}%
\providecommand \@@startlink[1]{}%
\providecommand \@@endlink[0]{}%
\providecommand \url  [0]{\begingroup\@sanitize@url \@url }%
\providecommand \@url [1]{\endgroup\@href {#1}{\urlprefix }}%
\providecommand \urlprefix  [0]{URL }%
\providecommand \Eprint [0]{\href }%
\providecommand \doibase [0]{https://doi.org/}%
\providecommand \selectlanguage [0]{\@gobble}%
\providecommand \bibinfo  [0]{\@secondoftwo}%
\providecommand \bibfield  [0]{\@secondoftwo}%
\providecommand \translation [1]{[#1]}%
\providecommand \BibitemOpen [0]{}%
\providecommand \bibitemStop [0]{}%
\providecommand \bibitemNoStop [0]{.\EOS\space}%
\providecommand \EOS [0]{\spacefactor3000\relax}%
\providecommand \BibitemShut  [1]{\csname bibitem#1\endcsname}%
\let\auto@bib@innerbib\@empty
%</preamble>
\bibitem [{\citenamefont {Jan}\ \emph {et~al.}(2015)\citenamefont {Jan},
  \citenamefont {Grimm}, \citenamefont {Tran}, \citenamefont {Lathrop},
  \citenamefont {Wollstein}, \citenamefont {Bilonick}, \citenamefont
  {Ishikawa}, \citenamefont {Kagemann}, \citenamefont {Schuman},\ and\
  \citenamefont {Sigal}}]{Jan:2015-4705:BOE}%
  \BibitemOpen
  \bibfield  {author} {\bibinfo {author} {\bibfnamefont {N.-J.}\ \bibnamefont
  {Jan}}, \bibinfo {author} {\bibfnamefont {J.~L.}\ \bibnamefont {Grimm}},
  \bibinfo {author} {\bibfnamefont {H.}~\bibnamefont {Tran}}, \bibinfo {author}
  {\bibfnamefont {K.~L.}\ \bibnamefont {Lathrop}}, \bibinfo {author}
  {\bibfnamefont {G.}~\bibnamefont {Wollstein}}, \bibinfo {author}
  {\bibfnamefont {R.~A.}\ \bibnamefont {Bilonick}}, \bibinfo {author}
  {\bibfnamefont {H.}~\bibnamefont {Ishikawa}}, \bibinfo {author}
  {\bibfnamefont {L.}~\bibnamefont {Kagemann}}, \bibinfo {author}
  {\bibfnamefont {J.~S.}\ \bibnamefont {Schuman}},\ and\ \bibinfo {author}
  {\bibfnamefont {I.~A.}\ \bibnamefont {Sigal}},\ }\bibfield  {title} {\bibinfo
  {title} {Polarization microscopy for characterizing fiber orientation of
  ocular tissues},\ }\href {https://doi.org/10.1364/BOE.6.004705} {\bibfield
  {journal} {\bibinfo  {journal} {Biomed. Opt. Express}\ }\textbf {\bibinfo
  {volume} {6}},\ \bibinfo {pages} {4705} (\bibinfo {year} {2015})}\BibitemShut
  {NoStop}%
\bibitem [{\citenamefont {Puthukkudy}\ \emph {et~al.}(2020)\citenamefont
  {Puthukkudy}, \citenamefont {Martins}, \citenamefont {Remer}, \citenamefont
  {Xu}, \citenamefont {Dubovik}, \citenamefont {Litvinov}, \citenamefont
  {McBride}, \citenamefont {Burton},\ and\ \citenamefont
  {Barbosa}}]{Puthukkudy:2020-5207:RAR}%
  \BibitemOpen
  \bibfield  {author} {\bibinfo {author} {\bibfnamefont {A.}~\bibnamefont
  {Puthukkudy}}, \bibinfo {author} {\bibfnamefont {J.~V.}\ \bibnamefont
  {Martins}}, \bibinfo {author} {\bibfnamefont {L.~A.}\ \bibnamefont {Remer}},
  \bibinfo {author} {\bibfnamefont {X.~G.}\ \bibnamefont {Xu}}, \bibinfo
  {author} {\bibfnamefont {O.}~\bibnamefont {Dubovik}}, \bibinfo {author}
  {\bibfnamefont {P.}~\bibnamefont {Litvinov}}, \bibinfo {author}
  {\bibfnamefont {B.}~\bibnamefont {McBride}}, \bibinfo {author} {\bibfnamefont
  {S.}~\bibnamefont {Burton}},\ and\ \bibinfo {author} {\bibfnamefont
  {H.~M.~J.}\ \bibnamefont {Barbosa}},\ }\bibfield  {title} {{\selectlanguage
  {English}\bibinfo {title} {Retrieval of aerosol properties from airborne
  hyper-angular rainbow polarimeter (airharp) observations during acepol
  2017}},\ }\href {https://doi.org/10.5194/amt-13-5207-2020} {\bibfield
  {journal} {\bibinfo  {journal} {Atmos. Meas. Tech.}\ }\textbf {\bibinfo
  {volume} {13}},\ \bibinfo {pages} {5207} (\bibinfo {year}
  {2020})}\BibitemShut {NoStop}%
\bibitem [{\citenamefont {Chekhova}\ and\ \citenamefont
  {Banzer}(2021)}]{Chekhova:2021:PolarizationLight}%
  \BibitemOpen
  \bibfield  {author} {\bibinfo {author} {\bibfnamefont {M.}~\bibnamefont
  {Chekhova}}\ and\ \bibinfo {author} {\bibfnamefont {P.}~\bibnamefont
  {Banzer}},\ }\href {https://doi.org/10.1515/9783110668025} {\emph {\bibinfo
  {title} {{Polarization of Light In Classical, Quantum, and Nonlinear
  Optics}}}}\ (\bibinfo  {publisher} {De Gruyter},\ \bibinfo {address}
  {Berlin},\ \bibinfo {year} {2021})\BibitemShut {NoStop}%
\bibitem [{\citenamefont {Martinez}(2018)}]{Martinez:2018-750:SCI}%
  \BibitemOpen
  \bibfield  {author} {\bibinfo {author} {\bibfnamefont {A.}~\bibnamefont
  {Martinez}},\ }\bibfield  {title} {{\selectlanguage {English}\bibinfo {title}
  {Polarimetry enabled by nanophotonics}},\ }\href
  {https://doi.org/10.1126/science.aau7494} {\bibfield  {journal} {\bibinfo
  {journal} {Science}\ }\textbf {\bibinfo {volume} {362}},\ \bibinfo {pages}
  {750} (\bibinfo {year} {2018})}\BibitemShut {NoStop}%
\bibitem [{\citenamefont {Rubin}\ \emph {et~al.}(2019)\citenamefont {Rubin},
  \citenamefont {D'Aversa}, \citenamefont {Chevalier}, \citenamefont {Shi},
  \citenamefont {Chen},\ and\ \citenamefont
  {Capasso}}]{Rubin:2019-eaax1839:SCI}%
  \BibitemOpen
  \bibfield  {author} {\bibinfo {author} {\bibfnamefont {N.~A.}\ \bibnamefont
  {Rubin}}, \bibinfo {author} {\bibfnamefont {G.}~\bibnamefont {D'Aversa}},
  \bibinfo {author} {\bibfnamefont {P.}~\bibnamefont {Chevalier}}, \bibinfo
  {author} {\bibfnamefont {Z.~J.}\ \bibnamefont {Shi}}, \bibinfo {author}
  {\bibfnamefont {W.~T.}\ \bibnamefont {Chen}},\ and\ \bibinfo {author}
  {\bibfnamefont {F.}~\bibnamefont {Capasso}},\ }\bibfield  {title}
  {{\selectlanguage {English}\bibinfo {title} {Matrix {F}ourier optics enables
  a compact full-{S}tokes polarization camera}},\ }\href
  {https://doi.org/10.1126/science.aax1839} {\bibfield  {journal} {\bibinfo
  {journal} {Science}\ }\textbf {\bibinfo {volume} {365}},\ \bibinfo {pages}
  {eaax1839} (\bibinfo {year} {2019})}\BibitemShut {NoStop}%
\bibitem [{\citenamefont {Stav}\ \emph {et~al.}(2018)\citenamefont {Stav},
  \citenamefont {Faerman}, \citenamefont {Maguid}, \citenamefont {Oren},
  \citenamefont {Kleiner}, \citenamefont {Hasman},\ and\ \citenamefont
  {Segev}}]{Stav:2018-1101:SCI}%
  \BibitemOpen
  \bibfield  {author} {\bibinfo {author} {\bibfnamefont {T.}~\bibnamefont
  {Stav}}, \bibinfo {author} {\bibfnamefont {A.}~\bibnamefont {Faerman}},
  \bibinfo {author} {\bibfnamefont {E.}~\bibnamefont {Maguid}}, \bibinfo
  {author} {\bibfnamefont {D.}~\bibnamefont {Oren}}, \bibinfo {author}
  {\bibfnamefont {V.}~\bibnamefont {Kleiner}}, \bibinfo {author} {\bibfnamefont
  {E.}~\bibnamefont {Hasman}},\ and\ \bibinfo {author} {\bibfnamefont
  {M.}~\bibnamefont {Segev}},\ }\bibfield  {title} {{\selectlanguage
  {English}\bibinfo {title} {Quantum entanglement of the spin and orbital
  angular momentum of photons using metamaterials}},\ }\href
  {https://doi.org/10.1126/science.aat9042} {\bibfield  {journal} {\bibinfo
  {journal} {Science}\ }\textbf {\bibinfo {volume} {361}},\ \bibinfo {pages}
  {1101} (\bibinfo {year} {2018})}\BibitemShut {NoStop}%
\bibitem [{\citenamefont {Wang}\ \emph {et~al.}(2018)\citenamefont {Wang},
  \citenamefont {Titchener}, \citenamefont {Kruk}, \citenamefont {Xu},
  \citenamefont {Chung}, \citenamefont {Parry}, \citenamefont {Kravchenko},
  \citenamefont {Chen}, \citenamefont {Solntsev}, \citenamefont {Kivshar},
  \citenamefont {Neshev},\ and\ \citenamefont
  {Sukhorukov}}]{Wang:2018-1104:SCI}%
  \BibitemOpen
  \bibfield  {author} {\bibinfo {author} {\bibfnamefont {K.}~\bibnamefont
  {Wang}}, \bibinfo {author} {\bibfnamefont {J.~G.}\ \bibnamefont {Titchener}},
  \bibinfo {author} {\bibfnamefont {S.~S.}\ \bibnamefont {Kruk}}, \bibinfo
  {author} {\bibfnamefont {L.}~\bibnamefont {Xu}}, \bibinfo {author}
  {\bibfnamefont {H.~P.}\ \bibnamefont {Chung}}, \bibinfo {author}
  {\bibfnamefont {M.}~\bibnamefont {Parry}}, \bibinfo {author} {\bibfnamefont
  {I.~I.}\ \bibnamefont {Kravchenko}}, \bibinfo {author} {\bibfnamefont
  {Y.~H.}\ \bibnamefont {Chen}}, \bibinfo {author} {\bibfnamefont {A.~S.}\
  \bibnamefont {Solntsev}}, \bibinfo {author} {\bibfnamefont {Y.~S.}\
  \bibnamefont {Kivshar}}, \bibinfo {author} {\bibfnamefont {D.~N.}\
  \bibnamefont {Neshev}},\ and\ \bibinfo {author} {\bibfnamefont {A.~A.}\
  \bibnamefont {Sukhorukov}},\ }\bibfield  {title} {{\selectlanguage
  {English}\bibinfo {title} {Quantum metasurface for multiphoton interference
  and state reconstruction}},\ }\href {https://doi.org/10.1126/science.aat8196}
  {\bibfield  {journal} {\bibinfo  {journal} {Science}\ }\textbf {\bibinfo
  {volume} {361}},\ \bibinfo {pages} {1104} (\bibinfo {year}
  {2018})}\BibitemShut {NoStop}%
\bibitem [{\citenamefont {Georgi}\ \emph {et~al.}(2019)\citenamefont {Georgi},
  \citenamefont {Massaro}, \citenamefont {Luo}, \citenamefont {Sain},
  \citenamefont {Montaut}, \citenamefont {Herrmann}, \citenamefont {Weiss},
  \citenamefont {Li}, \citenamefont {Silberhorn},\ and\ \citenamefont
  {Zentgraf}}]{Georgi:2019-70:LSA}%
  \BibitemOpen
  \bibfield  {author} {\bibinfo {author} {\bibfnamefont {P.}~\bibnamefont
  {Georgi}}, \bibinfo {author} {\bibfnamefont {M.}~\bibnamefont {Massaro}},
  \bibinfo {author} {\bibfnamefont {K.~H.}\ \bibnamefont {Luo}}, \bibinfo
  {author} {\bibfnamefont {B.}~\bibnamefont {Sain}}, \bibinfo {author}
  {\bibfnamefont {N.}~\bibnamefont {Montaut}}, \bibinfo {author} {\bibfnamefont
  {H.}~\bibnamefont {Herrmann}}, \bibinfo {author} {\bibfnamefont
  {T.}~\bibnamefont {Weiss}}, \bibinfo {author} {\bibfnamefont {G.~X.}\
  \bibnamefont {Li}}, \bibinfo {author} {\bibfnamefont {C.}~\bibnamefont
  {Silberhorn}},\ and\ \bibinfo {author} {\bibfnamefont {T.}~\bibnamefont
  {Zentgraf}},\ }\bibfield  {title} {{\selectlanguage {English}\bibinfo {title}
  {Metasurface interferometry toward quantum sensors}},\ }\href
  {https://doi.org/10.1038/s41377-019-0182-6} {\bibfield  {journal} {\bibinfo
  {journal} {Light Sci. Appl.}\ }\textbf {\bibinfo {volume} {8}},\ \bibinfo
  {pages} {70} (\bibinfo {year} {2019})}\BibitemShut {NoStop}%
\bibitem [{\citenamefont {Altuzarra}\ \emph {et~al.}(2019)\citenamefont
  {Altuzarra}, \citenamefont {Lyons}, \citenamefont {Yuan}, \citenamefont
  {Simpson}, \citenamefont {Roger}, \citenamefont {Ben-Benjamin},\ and\
  \citenamefont {Faccio}}]{Altuzarra:2019-20101:PRA}%
  \BibitemOpen
  \bibfield  {author} {\bibinfo {author} {\bibfnamefont {C.}~\bibnamefont
  {Altuzarra}}, \bibinfo {author} {\bibfnamefont {A.}~\bibnamefont {Lyons}},
  \bibinfo {author} {\bibfnamefont {G.~H.}\ \bibnamefont {Yuan}}, \bibinfo
  {author} {\bibfnamefont {C.}~\bibnamefont {Simpson}}, \bibinfo {author}
  {\bibfnamefont {T.}~\bibnamefont {Roger}}, \bibinfo {author} {\bibfnamefont
  {J.~S.}\ \bibnamefont {Ben-Benjamin}},\ and\ \bibinfo {author} {\bibfnamefont
  {D.}~\bibnamefont {Faccio}},\ }\bibfield  {title} {{\selectlanguage
  {English}\bibinfo {title} {Imaging of polarization-sensitive metasurfaces
  with quantum entanglement}},\ }\href
  {https://doi.org/10.1103/PhysRevA.99.020101} {\bibfield  {journal} {\bibinfo
  {journal} {Phys. Rev. A}\ }\textbf {\bibinfo {volume} {99}},\ \bibinfo
  {pages} {020101} (\bibinfo {year} {2019})}\BibitemShut {NoStop}%
\bibitem [{\citenamefont {Solntsev}\ \emph {et~al.}(2021)\citenamefont
  {Solntsev}, \citenamefont {Agarwal},\ and\ \citenamefont
  {Kivshar}}]{Solntsev:2021-327:NPHOT}%
  \BibitemOpen
  \bibfield  {author} {\bibinfo {author} {\bibfnamefont {A.~S.}\ \bibnamefont
  {Solntsev}}, \bibinfo {author} {\bibfnamefont {G.~S.}\ \bibnamefont
  {Agarwal}},\ and\ \bibinfo {author} {\bibfnamefont {Y.~S.}\ \bibnamefont
  {Kivshar}},\ }\bibfield  {title} {\bibinfo {title} {{Metasurfaces for Quantum
  Photonics}},\ }\href {https://doi.org/10.1038/s41566-021-00793-z} {\bibfield
  {journal} {\bibinfo  {journal} {Nat. Photon.}\ }\textbf {\bibinfo {volume}
  {15}},\ \bibinfo {pages} {327} (\bibinfo {year} {2021})}\BibitemShut
  {NoStop}%
\bibitem [{\citenamefont {Kellock}\ \emph {et~al.}(2014)\citenamefont
  {Kellock}, \citenamefont {Setala}, \citenamefont {Friberg},\ and\
  \citenamefont {Shirai}}]{Kellock:2014-55702:JOPT}%
  \BibitemOpen
  \bibfield  {author} {\bibinfo {author} {\bibfnamefont {H.}~\bibnamefont
  {Kellock}}, \bibinfo {author} {\bibfnamefont {T.}~\bibnamefont {Setala}},
  \bibinfo {author} {\bibfnamefont {A.~T.}\ \bibnamefont {Friberg}},\ and\
  \bibinfo {author} {\bibfnamefont {T.}~\bibnamefont {Shirai}},\ }\bibfield
  {title} {{\selectlanguage {English}\bibinfo {title} {Polarimetry by classical
  ghost diffraction}},\ }\href {https://doi.org/10.1088/2040-8978/16/5/055702}
  {\bibfield  {journal} {\bibinfo  {journal} {J. Opt.}\ }\textbf {\bibinfo
  {volume} {16}},\ \bibinfo {pages} {055702} (\bibinfo {year}
  {2014})}\BibitemShut {NoStop}%
\bibitem [{\citenamefont {Pittman}\ \emph {et~al.}(1995)\citenamefont
  {Pittman}, \citenamefont {Shih}, \citenamefont {Strekalov},\ and\
  \citenamefont {Sergienko}}]{Pittman:1995-3429:PRA}%
  \BibitemOpen
  \bibfield  {author} {\bibinfo {author} {\bibfnamefont {T.~B.}\ \bibnamefont
  {Pittman}}, \bibinfo {author} {\bibfnamefont {Y.~H.}\ \bibnamefont {Shih}},
  \bibinfo {author} {\bibfnamefont {D.~V.}\ \bibnamefont {Strekalov}},\ and\
  \bibinfo {author} {\bibfnamefont {A.~V.}\ \bibnamefont {Sergienko}},\
  }\bibfield  {title} {{\selectlanguage {English}\bibinfo {title} {Optical
  imaging by means of 2-photon quantum entanglement}},\ }\href
  {https://doi.org/10.1103/PhysRevA.52.R3429} {\bibfield  {journal} {\bibinfo
  {journal} {Phys. Rev. A}\ }\textbf {\bibinfo {volume} {52}},\ \bibinfo
  {pages} {R3429} (\bibinfo {year} {1995})}\BibitemShut {NoStop}%
\bibitem [{\citenamefont {Valencia}\ \emph {et~al.}(2005)\citenamefont
  {Valencia}, \citenamefont {Scarcelli}, \citenamefont {D'Angelo},\ and\
  \citenamefont {Shih}}]{Valencia:2005-63601:PRL}%
  \BibitemOpen
  \bibfield  {author} {\bibinfo {author} {\bibfnamefont {A.}~\bibnamefont
  {Valencia}}, \bibinfo {author} {\bibfnamefont {G.}~\bibnamefont {Scarcelli}},
  \bibinfo {author} {\bibfnamefont {M.}~\bibnamefont {D'Angelo}},\ and\
  \bibinfo {author} {\bibfnamefont {Y.}~\bibnamefont {Shih}},\ }\bibfield
  {title} {{\selectlanguage {English}\bibinfo {title} {Two-photon imaging with
  thermal light}},\ }\href {https://doi.org/10.1103/PhysRevLett.94.063601}
  {\bibfield  {journal} {\bibinfo  {journal} {Phys. Rev. Lett.}\ }\textbf
  {\bibinfo {volume} {94}},\ \bibinfo {pages} {063601} (\bibinfo {year}
  {2005})}\BibitemShut {NoStop}%
\bibitem [{\citenamefont {Chan}\ \emph {et~al.}(2009)\citenamefont {Chan},
  \citenamefont {O'Sullivan},\ and\ \citenamefont
  {Boyd}}]{Chan:2009-33808:PRA}%
  \BibitemOpen
  \bibfield  {author} {\bibinfo {author} {\bibfnamefont {K.~W.~C.}\
  \bibnamefont {Chan}}, \bibinfo {author} {\bibfnamefont {M.~N.}\ \bibnamefont
  {O'Sullivan}},\ and\ \bibinfo {author} {\bibfnamefont {R.~W.}\ \bibnamefont
  {Boyd}},\ }\bibfield  {title} {{\selectlanguage {English}\bibinfo {title}
  {Two-color ghost imaging}},\ }\href
  {https://doi.org/10.1103/PhysRevA.79.033808} {\bibfield  {journal} {\bibinfo
  {journal} {Phys. Rev. A}\ }\textbf {\bibinfo {volume} {79}},\ \bibinfo
  {pages} {033808} (\bibinfo {year} {2009})}\BibitemShut {NoStop}%
\bibitem [{\citenamefont {Karmakar}\ and\ \citenamefont
  {Shih}(2010)}]{Karmakar:2010-33845:PRA}%
  \BibitemOpen
  \bibfield  {author} {\bibinfo {author} {\bibfnamefont {S.}~\bibnamefont
  {Karmakar}}\ and\ \bibinfo {author} {\bibfnamefont {Y.~H.}\ \bibnamefont
  {Shih}},\ }\bibfield  {title} {{\selectlanguage {English}\bibinfo {title}
  {Two-color ghost imaging with enhanced angular resolving power}},\ }\href
  {https://doi.org/10.1103/PhysRevA.81.033845} {\bibfield  {journal} {\bibinfo
  {journal} {Phys. Rev. A}\ }\textbf {\bibinfo {volume} {81}},\ \bibinfo
  {pages} {033845} (\bibinfo {year} {2010})}\BibitemShut {NoStop}%
\bibitem [{\citenamefont {Aspden}\ \emph {et~al.}(2015)\citenamefont {Aspden},
  \citenamefont {Gemmell}, \citenamefont {Morris}, \citenamefont {Tasca},
  \citenamefont {Mertens}, \citenamefont {Tanner}, \citenamefont {Kirkwood},
  \citenamefont {Ruggeri}, \citenamefont {Tosi}, \citenamefont {Boyd},
  \citenamefont {Buller}, \citenamefont {Hadfield},\ and\ \citenamefont
  {Padgett}}]{Aspden:2015-1049:OPT}%
  \BibitemOpen
  \bibfield  {author} {\bibinfo {author} {\bibfnamefont {R.~S.}\ \bibnamefont
  {Aspden}}, \bibinfo {author} {\bibfnamefont {N.~R.}\ \bibnamefont {Gemmell}},
  \bibinfo {author} {\bibfnamefont {P.~A.}\ \bibnamefont {Morris}}, \bibinfo
  {author} {\bibfnamefont {D.~S.}\ \bibnamefont {Tasca}}, \bibinfo {author}
  {\bibfnamefont {L.}~\bibnamefont {Mertens}}, \bibinfo {author} {\bibfnamefont
  {M.~G.}\ \bibnamefont {Tanner}}, \bibinfo {author} {\bibfnamefont {R.~A.}\
  \bibnamefont {Kirkwood}}, \bibinfo {author} {\bibfnamefont {A.}~\bibnamefont
  {Ruggeri}}, \bibinfo {author} {\bibfnamefont {A.}~\bibnamefont {Tosi}},
  \bibinfo {author} {\bibfnamefont {R.~W.}\ \bibnamefont {Boyd}}, \bibinfo
  {author} {\bibfnamefont {G.~S.}\ \bibnamefont {Buller}}, \bibinfo {author}
  {\bibfnamefont {R.~H.}\ \bibnamefont {Hadfield}},\ and\ \bibinfo {author}
  {\bibfnamefont {M.~J.}\ \bibnamefont {Padgett}},\ }\bibfield  {title}
  {{\selectlanguage {English}\bibinfo {title} {Photon-sparse microscopy:
  visible light imaging using infrared illumination}},\ }\href
  {https://doi.org/10.1364/OPTICA.2.001049} {\bibfield  {journal} {\bibinfo
  {journal} {Optica}\ }\textbf {\bibinfo {volume} {2}},\ \bibinfo {pages}
  {1049} (\bibinfo {year} {2015})}\BibitemShut {NoStop}%
\bibitem [{\citenamefont {Erkmen}\ and\ \citenamefont
  {Shapiro}(2010)}]{Erkmen:2010-405:ADOP}%
  \BibitemOpen
  \bibfield  {author} {\bibinfo {author} {\bibfnamefont {B.~I.}\ \bibnamefont
  {Erkmen}}\ and\ \bibinfo {author} {\bibfnamefont {J.~H.}\ \bibnamefont
  {Shapiro}},\ }\bibfield  {title} {{\selectlanguage {English}\bibinfo {title}
  {Ghost imaging: from quantum to classical to computational}},\ }\href
  {https://doi.org/10.1364/AOP.2.000405} {\bibfield  {journal} {\bibinfo
  {journal} {Adv. Opt. Photon.}\ }\textbf {\bibinfo {volume} {2}},\ \bibinfo
  {pages} {405} (\bibinfo {year} {2010})}\BibitemShut {NoStop}%
\bibitem [{\citenamefont {Brida}\ \emph {et~al.}(2010)\citenamefont {Brida},
  \citenamefont {Genovese},\ and\ \citenamefont
  {Berchera}}]{Brida:2010-227:NPHOT}%
  \BibitemOpen
  \bibfield  {author} {\bibinfo {author} {\bibfnamefont {G.}~\bibnamefont
  {Brida}}, \bibinfo {author} {\bibfnamefont {M.}~\bibnamefont {Genovese}},\
  and\ \bibinfo {author} {\bibfnamefont {I.~R.}\ \bibnamefont {Berchera}},\
  }\bibfield  {title} {{\selectlanguage {English}\bibinfo {title} {Experimental
  realization of sub-shot-noise quantum imaging}},\ }\href
  {https://doi.org/10.1038/NPHOTON.2010.29} {\bibfield  {journal} {\bibinfo
  {journal} {Nat. Photon.}\ }\textbf {\bibinfo {volume} {4}},\ \bibinfo {pages}
  {227} (\bibinfo {year} {2010})}\BibitemShut {NoStop}%
\bibitem [{\citenamefont {Morris}\ \emph {et~al.}(2015)\citenamefont {Morris},
  \citenamefont {Aspden}, \citenamefont {Bell}, \citenamefont {Boyd},\ and\
  \citenamefont {Padgett}}]{Morris:2015-5913:NCOM}%
  \BibitemOpen
  \bibfield  {author} {\bibinfo {author} {\bibfnamefont {P.~A.}\ \bibnamefont
  {Morris}}, \bibinfo {author} {\bibfnamefont {R.~S.}\ \bibnamefont {Aspden}},
  \bibinfo {author} {\bibfnamefont {J.~E.~C.}\ \bibnamefont {Bell}}, \bibinfo
  {author} {\bibfnamefont {R.~W.}\ \bibnamefont {Boyd}},\ and\ \bibinfo
  {author} {\bibfnamefont {M.~J.}\ \bibnamefont {Padgett}},\ }\bibfield
  {title} {{\selectlanguage {English}\bibinfo {title} {Imaging with a small
  number of photons}},\ }\href {https://doi.org/10.1038/ncomms6913} {\bibfield
  {journal} {\bibinfo  {journal} {Nat. Commun.}\ }\textbf {\bibinfo {volume}
  {6}},\ \bibinfo {pages} {5913} (\bibinfo {year} {2015})}\BibitemShut
  {NoStop}%
\bibitem [{\citenamefont {Hannonen}\ \emph {et~al.}(2016)\citenamefont
  {Hannonen}, \citenamefont {Friberg},\ and\ \citenamefont
  {Setala}}]{Hannonen:2016-4943:OL}%
  \BibitemOpen
  \bibfield  {author} {\bibinfo {author} {\bibfnamefont {A.}~\bibnamefont
  {Hannonen}}, \bibinfo {author} {\bibfnamefont {A.~T.}\ \bibnamefont
  {Friberg}},\ and\ \bibinfo {author} {\bibfnamefont {T.}~\bibnamefont
  {Setala}},\ }\bibfield  {title} {{\selectlanguage {English}\bibinfo {title}
  {Classical spectral ghost ellipsometry}},\ }\href
  {https://doi.org/10.1364/OL.41.004943} {\bibfield  {journal} {\bibinfo
  {journal} {Opt. Lett.}\ }\textbf {\bibinfo {volume} {41}},\ \bibinfo {pages}
  {4943} (\bibinfo {year} {2016})}\BibitemShut {NoStop}%
\bibitem [{\citenamefont {Hannonen}\ \emph {et~al.}(2017)\citenamefont
  {Hannonen}, \citenamefont {Friberg},\ and\ \citenamefont
  {Setala}}]{Hannonen:2017-1360:JOSA}%
  \BibitemOpen
  \bibfield  {author} {\bibinfo {author} {\bibfnamefont {A.}~\bibnamefont
  {Hannonen}}, \bibinfo {author} {\bibfnamefont {A.~T.}\ \bibnamefont
  {Friberg}},\ and\ \bibinfo {author} {\bibfnamefont {T.}~\bibnamefont
  {Setala}},\ }\bibfield  {title} {{\selectlanguage {English}\bibinfo {title}
  {Classical ghost-imaging spectral ellipsometer}},\ }\href
  {https://doi.org/10.1364/JOSAA.34.001360} {\bibfield  {journal} {\bibinfo
  {journal} {J. Opt. Soc. Am. A}\ }\textbf {\bibinfo {volume} {34}},\ \bibinfo
  {pages} {1360} (\bibinfo {year} {2017})}\BibitemShut {NoStop}%
\bibitem [{\citenamefont {Janassek}\ \emph {et~al.}(2018)\citenamefont
  {Janassek}, \citenamefont {Blumenstein},\ and\ \citenamefont
  {Elsasser}}]{Janassek:2018-883:OL}%
  \BibitemOpen
  \bibfield  {author} {\bibinfo {author} {\bibfnamefont {P.}~\bibnamefont
  {Janassek}}, \bibinfo {author} {\bibfnamefont {S.}~\bibnamefont
  {Blumenstein}},\ and\ \bibinfo {author} {\bibfnamefont {W.}~\bibnamefont
  {Elsasser}},\ }\bibfield  {title} {{\selectlanguage {English}\bibinfo {title}
  {Recovering a hidden polarization by ghost polarimetry}},\ }\href
  {https://doi.org/10.1364/OL.43.000883} {\bibfield  {journal} {\bibinfo
  {journal} {Opt. Lett.}\ }\textbf {\bibinfo {volume} {43}},\ \bibinfo {pages}
  {883} (\bibinfo {year} {2018})}\BibitemShut {NoStop}%
\bibitem [{\citenamefont {Shi}\ \emph {et~al.}(2018)\citenamefont {Shi},
  \citenamefont {Zhang}, \citenamefont {Huang}, \citenamefont {Wang},
  \citenamefont {Yuan}, \citenamefont {Cao}, \citenamefont {Xie}, \citenamefont
  {Liu},\ and\ \citenamefont {Zhu}}]{Shi:2018-100:OLE}%
  \BibitemOpen
  \bibfield  {author} {\bibinfo {author} {\bibfnamefont {D.~F.}\ \bibnamefont
  {Shi}}, \bibinfo {author} {\bibfnamefont {J.~M.}\ \bibnamefont {Zhang}},
  \bibinfo {author} {\bibfnamefont {J.}~\bibnamefont {Huang}}, \bibinfo
  {author} {\bibfnamefont {Y.~J.}\ \bibnamefont {Wang}}, \bibinfo {author}
  {\bibfnamefont {K.}~\bibnamefont {Yuan}}, \bibinfo {author} {\bibfnamefont
  {K.~F.}\ \bibnamefont {Cao}}, \bibinfo {author} {\bibfnamefont {C.~B.}\
  \bibnamefont {Xie}}, \bibinfo {author} {\bibfnamefont {D.}~\bibnamefont
  {Liu}},\ and\ \bibinfo {author} {\bibfnamefont {W.~Y.}\ \bibnamefont {Zhu}},\
  }\bibfield  {title} {{\selectlanguage {English}\bibinfo {title}
  {Polarization-multiplexing ghost imaging}},\ }\href
  {https://doi.org/10.1016/j.optlaseng.2017.10.022} {\bibfield  {journal}
  {\bibinfo  {journal} {Opt. Lasers Eng.}\ }\textbf {\bibinfo {volume} {102}},\
  \bibinfo {pages} {100} (\bibinfo {year} {2018})}\BibitemShut {NoStop}%
\bibitem [{\citenamefont {Chirkin}\ \emph {et~al.}(2018)\citenamefont
  {Chirkin}, \citenamefont {Gostev}, \citenamefont {Agapov},\ and\
  \citenamefont {Magnitskiy}}]{Chirkin:2018-115404:LPL}%
  \BibitemOpen
  \bibfield  {author} {\bibinfo {author} {\bibfnamefont {A.~S.}\ \bibnamefont
  {Chirkin}}, \bibinfo {author} {\bibfnamefont {P.~P.}\ \bibnamefont {Gostev}},
  \bibinfo {author} {\bibfnamefont {D.~P.}\ \bibnamefont {Agapov}},\ and\
  \bibinfo {author} {\bibfnamefont {S.~A.}\ \bibnamefont {Magnitskiy}},\
  }\bibfield  {title} {{\selectlanguage {English}\bibinfo {title} {Ghost
  polarimetry: ghost imaging of polarization-sensitive objects}},\ }\href
  {https://doi.org/10.1088/1612-202X/aae4a6} {\bibfield  {journal} {\bibinfo
  {journal} {Laser Phys. Lett.}\ }\textbf {\bibinfo {volume} {15}},\ \bibinfo
  {pages} {115404} (\bibinfo {year} {2018})}\BibitemShut {NoStop}%
\bibitem [{\citenamefont {Hannonen}\ \emph {et~al.}(2020)\citenamefont
  {Hannonen}, \citenamefont {Hoenders}, \citenamefont {Elsasser}, \citenamefont
  {Friberg},\ and\ \citenamefont {Setala}}]{Hannonen:2020-714:JOSA}%
  \BibitemOpen
  \bibfield  {author} {\bibinfo {author} {\bibfnamefont {A.}~\bibnamefont
  {Hannonen}}, \bibinfo {author} {\bibfnamefont {B.~J.}\ \bibnamefont
  {Hoenders}}, \bibinfo {author} {\bibfnamefont {W.}~\bibnamefont {Elsasser}},
  \bibinfo {author} {\bibfnamefont {A.~T.}\ \bibnamefont {Friberg}},\ and\
  \bibinfo {author} {\bibfnamefont {T.}~\bibnamefont {Setala}},\ }\bibfield
  {title} {{\selectlanguage {English}\bibinfo {title} {Ghost polarimetry using
  {S}tokes correlations}},\ }\href {https://doi.org/10.1364/JOSAA.385851}
  {\bibfield  {journal} {\bibinfo  {journal} {J. Opt. Soc. Am. A}\ }\textbf
  {\bibinfo {volume} {37}},\ \bibinfo {pages} {714} (\bibinfo {year}
  {2020})}\BibitemShut {NoStop}%
\bibitem [{\citenamefont {Rosskopf}\ \emph {et~al.}(2020)\citenamefont
  {Rosskopf}, \citenamefont {Mohr},\ and\ \citenamefont
  {Elsasser}}]{Rosskopf:2020-34062:PRAP}%
  \BibitemOpen
  \bibfield  {author} {\bibinfo {author} {\bibfnamefont {M.}~\bibnamefont
  {Rosskopf}}, \bibinfo {author} {\bibfnamefont {T.}~\bibnamefont {Mohr}},\
  and\ \bibinfo {author} {\bibfnamefont {W.}~\bibnamefont {Elsasser}},\
  }\bibfield  {title} {{\selectlanguage {English}\bibinfo {title} {Ghost
  polarization communication}},\ }\href
  {https://doi.org/10.1103/PhysRevApplied.13.034062} {\bibfield  {journal}
  {\bibinfo  {journal} {Phys. Rev. Appl.}\ }\textbf {\bibinfo {volume} {13}},\
  \bibinfo {pages} {034062} (\bibinfo {year} {2020})}\BibitemShut {NoStop}%
\bibitem [{\citenamefont {Magnitskiy}\ \emph {et~al.}(2020)\citenamefont
  {Magnitskiy}, \citenamefont {Agapov},\ and\ \citenamefont
  {Chirkin}}]{Magnitskiy:2020-3641:OL}%
  \BibitemOpen
  \bibfield  {author} {\bibinfo {author} {\bibfnamefont {S.}~\bibnamefont
  {Magnitskiy}}, \bibinfo {author} {\bibfnamefont {D.}~\bibnamefont {Agapov}},\
  and\ \bibinfo {author} {\bibfnamefont {A.}~\bibnamefont {Chirkin}},\
  }\bibfield  {title} {{\selectlanguage {English}\bibinfo {title} {Ghost
  polarimetry with unpolarized pseudo-thermal light}},\ }\href
  {https://doi.org/10.1364/OL.387234} {\bibfield  {journal} {\bibinfo
  {journal} {Opt. Lett.}\ }\textbf {\bibinfo {volume} {45}},\ \bibinfo {pages}
  {3641} (\bibinfo {year} {2020})}\BibitemShut {NoStop}%
\bibitem [{\citenamefont {Liu}\ \emph {et~al.}(2017)\citenamefont {Liu},
  \citenamefont {Yang}, \citenamefont {Guo}, \citenamefont {Shi}, \citenamefont
  {Guan}, \citenamefont {Zheng}, \citenamefont {Muhlenbernd}, \citenamefont
  {Li}, \citenamefont {Zentgraf},\ and\ \citenamefont
  {Zhang}}]{Liu:2017-e1701477:SCA}%
  \BibitemOpen
  \bibfield  {author} {\bibinfo {author} {\bibfnamefont {H.~C.}\ \bibnamefont
  {Liu}}, \bibinfo {author} {\bibfnamefont {B.~A.}\ \bibnamefont {Yang}},
  \bibinfo {author} {\bibfnamefont {Q.~H.}\ \bibnamefont {Guo}}, \bibinfo
  {author} {\bibfnamefont {J.~H.}\ \bibnamefont {Shi}}, \bibinfo {author}
  {\bibfnamefont {C.~Y.}\ \bibnamefont {Guan}}, \bibinfo {author}
  {\bibfnamefont {G.~X.}\ \bibnamefont {Zheng}}, \bibinfo {author}
  {\bibfnamefont {H.}~\bibnamefont {Muhlenbernd}}, \bibinfo {author}
  {\bibfnamefont {G.~X.}\ \bibnamefont {Li}}, \bibinfo {author} {\bibfnamefont
  {T.}~\bibnamefont {Zentgraf}},\ and\ \bibinfo {author} {\bibfnamefont
  {S.}~\bibnamefont {Zhang}},\ }\bibfield  {title} {{\selectlanguage
  {English}\bibinfo {title} {Single-pixel computational ghost imaging with
  helicity-dependent metasurface hologram}},\ }\href
  {https://doi.org/10.1126/sciadv.1701477} {\bibfield  {journal} {\bibinfo
  {journal} {Sci. Adv.}\ }\textbf {\bibinfo {volume} {3}},\ \bibinfo {pages}
  {e1701477} (\bibinfo {year} {2017})}\BibitemShut {NoStop}%
\bibitem [{\citenamefont {Malik}\ \emph {et~al.}(2010)\citenamefont {Malik},
  \citenamefont {Shin}, \citenamefont {O'Sullivan}, \citenamefont {Zerom},\
  and\ \citenamefont {Boyd}}]{Malik:2010-163602:PRL}%
  \BibitemOpen
  \bibfield  {author} {\bibinfo {author} {\bibfnamefont {M.}~\bibnamefont
  {Malik}}, \bibinfo {author} {\bibfnamefont {H.}~\bibnamefont {Shin}},
  \bibinfo {author} {\bibfnamefont {M.}~\bibnamefont {O'Sullivan}}, \bibinfo
  {author} {\bibfnamefont {P.}~\bibnamefont {Zerom}},\ and\ \bibinfo {author}
  {\bibfnamefont {R.~W.}\ \bibnamefont {Boyd}},\ }\bibfield  {title}
  {{\selectlanguage {English}\bibinfo {title} {Quantum ghost image
  identification with correlated photon pairs}},\ }\href
  {https://doi.org/10.1103/PhysRevLett.104.163602} {\bibfield  {journal}
  {\bibinfo  {journal} {Phys. Rev. Lett.}\ }\textbf {\bibinfo {volume} {104}},\
  \bibinfo {pages} {163602} (\bibinfo {year} {2010})}\BibitemShut {NoStop}%
\bibitem [{\citenamefont {Slussarenko}\ and\ \citenamefont
  {Pryde}(2019)}]{Slussarenko:2019-41303:APR}%
  \BibitemOpen
  \bibfield  {author} {\bibinfo {author} {\bibfnamefont {S.}~\bibnamefont
  {Slussarenko}}\ and\ \bibinfo {author} {\bibfnamefont {G.~J.}\ \bibnamefont
  {Pryde}},\ }\bibfield  {title} {{\selectlanguage {English}\bibinfo {title}
  {Photonic quantum information processing: A concise review}},\ }\href
  {https://doi.org/10.1063/1.5115814} {\bibfield  {journal} {\bibinfo
  {journal} {Appl. Phys. Rev.}\ }\textbf {\bibinfo {volume} {6}},\ \bibinfo
  {pages} {041303} (\bibinfo {year} {2019})}\BibitemShut {NoStop}%
\bibitem [{\citenamefont {Huang}\ \emph {et~al.}(2020)\citenamefont {Huang},
  \citenamefont {Wu}, \citenamefont {Fan},\ and\ \citenamefont
  {Zhu}}]{Huang:2020-180501:SCIS}%
  \BibitemOpen
  \bibfield  {author} {\bibinfo {author} {\bibfnamefont {H.~L.}\ \bibnamefont
  {Huang}}, \bibinfo {author} {\bibfnamefont {D.~C.}\ \bibnamefont {Wu}},
  \bibinfo {author} {\bibfnamefont {D.~J.}\ \bibnamefont {Fan}},\ and\ \bibinfo
  {author} {\bibfnamefont {X.~B.}\ \bibnamefont {Zhu}},\ }\bibfield  {title}
  {{\selectlanguage {English}\bibinfo {title} {Superconducting quantum
  computing: a review}},\ }\href {https://doi.org/10.1007/s11432-020-2881-9}
  {\bibfield  {journal} {\bibinfo  {journal} {Sci. China-Inf. Sci.}\ }\textbf
  {\bibinfo {volume} {63}},\ \bibinfo {pages} {180501} (\bibinfo {year}
  {2020})}\BibitemShut {NoStop}%
\bibitem [{\citenamefont {Gisin}\ and\ \citenamefont
  {Thew}(2007)}]{Gisin:2007-165:NPHOT}%
  \BibitemOpen
  \bibfield  {author} {\bibinfo {author} {\bibfnamefont {N.}~\bibnamefont
  {Gisin}}\ and\ \bibinfo {author} {\bibfnamefont {R.}~\bibnamefont {Thew}},\
  }\bibfield  {title} {{\selectlanguage {English}\bibinfo {title} {Quantum
  communication}},\ }\href {https://doi.org/10.1038/nphoton.2007.22} {\bibfield
   {journal} {\bibinfo  {journal} {Nat. Photon.}\ }\textbf {\bibinfo {volume}
  {1}},\ \bibinfo {pages} {165} (\bibinfo {year} {2007})}\BibitemShut {NoStop}%
\bibitem [{\citenamefont {Giovannetti}\ \emph {et~al.}(2011)\citenamefont
  {Giovannetti}, \citenamefont {Lloyd},\ and\ \citenamefont
  {Maccone}}]{Giovannetti:2011-222:NPHOT}%
  \BibitemOpen
  \bibfield  {author} {\bibinfo {author} {\bibfnamefont {V.}~\bibnamefont
  {Giovannetti}}, \bibinfo {author} {\bibfnamefont {S.}~\bibnamefont {Lloyd}},\
  and\ \bibinfo {author} {\bibfnamefont {L.}~\bibnamefont {Maccone}},\
  }\bibfield  {title} {{\selectlanguage {English}\bibinfo {title} {Advances in
  quantum metrology}},\ }\href {https://doi.org/10.1038/NPHOTON.2011.35}
  {\bibfield  {journal} {\bibinfo  {journal} {Nat. Photon.}\ }\textbf {\bibinfo
  {volume} {5}},\ \bibinfo {pages} {222} (\bibinfo {year} {2011})}\BibitemShut
  {NoStop}%
\bibitem [{\citenamefont {Pezze}\ \emph {et~al.}(2018)\citenamefont {Pezze},
  \citenamefont {Smerzi}, \citenamefont {Oberthaler}, \citenamefont {Schmied},\
  and\ \citenamefont {Treutlein}}]{Pezze:2018-35005:RMP}%
  \BibitemOpen
  \bibfield  {author} {\bibinfo {author} {\bibfnamefont {L.}~\bibnamefont
  {Pezze}}, \bibinfo {author} {\bibfnamefont {A.}~\bibnamefont {Smerzi}},
  \bibinfo {author} {\bibfnamefont {M.~K.}\ \bibnamefont {Oberthaler}},
  \bibinfo {author} {\bibfnamefont {R.}~\bibnamefont {Schmied}},\ and\ \bibinfo
  {author} {\bibfnamefont {P.}~\bibnamefont {Treutlein}},\ }\bibfield  {title}
  {{\selectlanguage {English}\bibinfo {title} {Quantum metrology with
  nonclassical states of atomic ensembles}},\ }\href
  {https://doi.org/10.1103/RevModPhys.90.035005} {\bibfield  {journal}
  {\bibinfo  {journal} {Rev. Mod. Phys.}\ }\textbf {\bibinfo {volume} {90}},\
  \bibinfo {pages} {035005} (\bibinfo {year} {2018})}\BibitemShut {NoStop}%
\bibitem [{\citenamefont {Shapiro}\ \emph {et~al.}(2015)\citenamefont
  {Shapiro}, \citenamefont {Venkatraman},\ and\ \citenamefont
  {Wong}}]{Shapiro:2015-10329:SRP}%
  \BibitemOpen
  \bibfield  {author} {\bibinfo {author} {\bibfnamefont {J.~H.}\ \bibnamefont
  {Shapiro}}, \bibinfo {author} {\bibfnamefont {D.}~\bibnamefont
  {Venkatraman}},\ and\ \bibinfo {author} {\bibfnamefont {F.~N.~C.}\
  \bibnamefont {Wong}},\ }\bibfield  {title} {{\selectlanguage
  {English}\bibinfo {title} {Classical imaging with undetected photons}},\
  }\href {https://doi.org/10.1038/srep10329} {\bibfield  {journal} {\bibinfo
  {journal} {Sci. Rep.}\ }\textbf {\bibinfo {volume} {5}},\ \bibinfo {pages}
  {10329} (\bibinfo {year} {2015})}\BibitemShut {NoStop}%
\bibitem [{\citenamefont {Bennink}\ \emph {et~al.}(2002)\citenamefont
  {Bennink}, \citenamefont {Bentley},\ and\ \citenamefont
  {Boyd}}]{Bennink:2002-113601:PRL}%
  \BibitemOpen
  \bibfield  {author} {\bibinfo {author} {\bibfnamefont {R.~S.}\ \bibnamefont
  {Bennink}}, \bibinfo {author} {\bibfnamefont {S.~J.}\ \bibnamefont
  {Bentley}},\ and\ \bibinfo {author} {\bibfnamefont {R.~W.}\ \bibnamefont
  {Boyd}},\ }\bibfield  {title} {{\selectlanguage {English}\bibinfo {title}
  {"two-photon" coincidence imaging with a classical source}},\ }\href
  {https://doi.org/10.1103/PhysRevLett.89.113601} {\bibfield  {journal}
  {\bibinfo  {journal} {Phys. Rev. Lett.}\ }\textbf {\bibinfo {volume} {89}},\
  \bibinfo {pages} {113601} (\bibinfo {year} {2002})}\BibitemShut {NoStop}%
\bibitem [{\citenamefont {Gatti}\ \emph {et~al.}(2004)\citenamefont {Gatti},
  \citenamefont {Brambilla}, \citenamefont {Bache},\ and\ \citenamefont
  {Lugiato}}]{Gatti:2004-93602:PRL}%
  \BibitemOpen
  \bibfield  {author} {\bibinfo {author} {\bibfnamefont {A.}~\bibnamefont
  {Gatti}}, \bibinfo {author} {\bibfnamefont {E.}~\bibnamefont {Brambilla}},
  \bibinfo {author} {\bibfnamefont {M.}~\bibnamefont {Bache}},\ and\ \bibinfo
  {author} {\bibfnamefont {L.~A.}\ \bibnamefont {Lugiato}},\ }\bibfield
  {title} {{\selectlanguage {English}\bibinfo {title} {Ghost imaging with
  thermal light: Comparing entanglement and classical correlation}},\ }\href
  {https://doi.org/10.1103/PhysRevLett.93.093602} {\bibfield  {journal}
  {\bibinfo  {journal} {Phys. Rev. Lett.}\ }\textbf {\bibinfo {volume} {93}},\
  \bibinfo {pages} {093602} (\bibinfo {year} {2004})}\BibitemShut {NoStop}%
\bibitem [{\citenamefont {Wootters}(1998)}]{Wootters:1998-2245:PRL}%
  \BibitemOpen
  \bibfield  {author} {\bibinfo {author} {\bibfnamefont {W.~K.}\ \bibnamefont
  {Wootters}},\ }\bibfield  {title} {{\selectlanguage {English}\bibinfo {title}
  {Entanglement of formation of an arbitrary state of two qubits}},\ }\href
  {https://doi.org/10.1103/PhysRevLett.80.2245} {\bibfield  {journal} {\bibinfo
   {journal} {Phys. Rev. Lett.}\ }\textbf {\bibinfo {volume} {80}},\ \bibinfo
  {pages} {2245} (\bibinfo {year} {1998})}\BibitemShut {NoStop}%
\bibitem [{\citenamefont {Aspect}\ \emph {et~al.}(1982)\citenamefont {Aspect},
  \citenamefont {Grangier},\ and\ \citenamefont {Roger}}]{Aspect:1982-91:PRL}%
  \BibitemOpen
  \bibfield  {author} {\bibinfo {author} {\bibfnamefont {A.}~\bibnamefont
  {Aspect}}, \bibinfo {author} {\bibfnamefont {P.}~\bibnamefont {Grangier}},\
  and\ \bibinfo {author} {\bibfnamefont {G.}~\bibnamefont {Roger}},\ }\bibfield
   {title} {{\selectlanguage {English}\bibinfo {title} {Experimental
  realization of einstein-podolsky-rosen-bohm gedankenexperiment - a new
  violation of {B}ell inequalities}},\ }\href
  {https://doi.org/10.1103/PhysRevLett.49.91} {\bibfield  {journal} {\bibinfo
  {journal} {Phys. Rev. Lett.}\ }\textbf {\bibinfo {volume} {49}},\ \bibinfo
  {pages} {91} (\bibinfo {year} {1982})}\BibitemShut {NoStop}%
\bibitem [{\citenamefont {Peters}\ \emph {et~al.}(2005)\citenamefont {Peters},
  \citenamefont {Barreiro}, \citenamefont {Goggin}, \citenamefont {Wei},\ and\
  \citenamefont {Kwiat}}]{Peters:2005-150502:PRL}%
  \BibitemOpen
  \bibfield  {author} {\bibinfo {author} {\bibfnamefont {N.~A.}\ \bibnamefont
  {Peters}}, \bibinfo {author} {\bibfnamefont {J.~T.}\ \bibnamefont
  {Barreiro}}, \bibinfo {author} {\bibfnamefont {M.~E.}\ \bibnamefont
  {Goggin}}, \bibinfo {author} {\bibfnamefont {T.~C.}\ \bibnamefont {Wei}},\
  and\ \bibinfo {author} {\bibfnamefont {P.~G.}\ \bibnamefont {Kwiat}},\
  }\bibfield  {title} {{\selectlanguage {English}\bibinfo {title} {Remote state
  preparation: Arbitrary remote control of photon polarization}},\ }\href
  {https://doi.org/10.1103/PhysRevLett.94.150502} {\bibfield  {journal}
  {\bibinfo  {journal} {Phys. Rev. Lett.}\ }\textbf {\bibinfo {volume} {94}},\
  \bibinfo {pages} {150502} (\bibinfo {year} {2005})}\BibitemShut {NoStop}%
\bibitem [{\citenamefont {Goldberg}\ \emph {et~al.}(2021)\citenamefont
  {Goldberg}, \citenamefont {De~La~Hoz}, \citenamefont {Bjork}, \citenamefont
  {Klimov}, \citenamefont {Grassl}, \citenamefont {Leuchs},\ and\ \citenamefont
  {Sanchez-Soro}}]{Goldberg:2021-1:ADOP}%
  \BibitemOpen
  \bibfield  {author} {\bibinfo {author} {\bibfnamefont {A.~Z.}\ \bibnamefont
  {Goldberg}}, \bibinfo {author} {\bibfnamefont {P.}~\bibnamefont {De~La~Hoz}},
  \bibinfo {author} {\bibfnamefont {G.}~\bibnamefont {Bjork}}, \bibinfo
  {author} {\bibfnamefont {A.~B.}\ \bibnamefont {Klimov}}, \bibinfo {author}
  {\bibfnamefont {M.}~\bibnamefont {Grassl}}, \bibinfo {author} {\bibfnamefont
  {G.}~\bibnamefont {Leuchs}},\ and\ \bibinfo {author} {\bibfnamefont {L.~L.}\
  \bibnamefont {Sanchez-Soro}},\ }\bibfield  {title} {{\selectlanguage
  {English}\bibinfo {title} {Quantum concepts in optical polarization}},\
  }\href {https://doi.org/10.1364/AOP.404175} {\bibfield  {journal} {\bibinfo
  {journal} {Adv. Opt. Photon.}\ }\textbf {\bibinfo {volume} {13}},\ \bibinfo
  {pages} {1} (\bibinfo {year} {2021})}\BibitemShut {NoStop}%
\bibitem [{\citenamefont {Lung}\ \emph {et~al.}(2020)\citenamefont {Lung},
  \citenamefont {Wang}, \citenamefont {Kamali}, \citenamefont {Zhang},
  \citenamefont {Rahmani}, \citenamefont {Neshev},\ and\ \citenamefont
  {Sukhorukov}}]{Lung:2020-3015:ACSP}%
  \BibitemOpen
  \bibfield  {author} {\bibinfo {author} {\bibfnamefont {S.}~\bibnamefont
  {Lung}}, \bibinfo {author} {\bibfnamefont {K.}~\bibnamefont {Wang}}, \bibinfo
  {author} {\bibfnamefont {K.~Z.}\ \bibnamefont {Kamali}}, \bibinfo {author}
  {\bibfnamefont {J.~H.}\ \bibnamefont {Zhang}}, \bibinfo {author}
  {\bibfnamefont {M.}~\bibnamefont {Rahmani}}, \bibinfo {author} {\bibfnamefont
  {D.~N.}\ \bibnamefont {Neshev}},\ and\ \bibinfo {author} {\bibfnamefont
  {A.~A.}\ \bibnamefont {Sukhorukov}},\ }\bibfield  {title} {{\selectlanguage
  {English}\bibinfo {title} {Complex-birefringent dielectric metasurfaces for
  arbitrary polarization-pair transformations}},\ }\href
  {https://doi.org/10.1021/acsphotonics.0c01044} {\bibfield  {journal}
  {\bibinfo  {journal} {ACS Photonics}\ }\textbf {\bibinfo {volume} {7}},\
  \bibinfo {pages} {3015} (\bibinfo {year} {2020})}\BibitemShut {NoStop}%
\end{thebibliography}%

\end{document}